\documentclass[useAMS,usenatbib,usegraphicx]{mn2e}
 
\usepackage{natbib}
\usepackage{amssymb}
\usepackage{amsfonts}
\usepackage{amsmath}
\usepackage{color}
\input epsf
 
\newcommand{\ltsimeq}{\raisebox{-0.6ex}{$\,\stackrel
        {\raisebox{-.2ex}{$\textstyle <$}}{\sim}\,$}}
\newcommand{\gtsimeq}{\raisebox{-0.6ex}{$\,\stackrel
        {\raisebox{-.2ex}{$\textstyle >$}}{\sim}\,$}}

\begin{document}

\title[Dark Energy with future HI surveys]{Forecasts for Dark Energy
  Measurements with Future HI Surveys}

\author[F. B. Abdalla et al.]{Filipe B. Abdalla$^1$, Chris Blake$^2$,
  Steve Rawlings$^3$ \\ \\ $^1$ Department of Physics \& Astronomy,
  University College London, Gower Street, London, WC1E 6BT,
  U.K. \\ $^2$ Centre for Astrophysics \& Supercomputing, Swinburne
  University of Technology, Hawthorn, VIC 3122, Australia\\ $^3$
  Department of Astrophysics, University of Oxford, Denys Wilkinson
  Building, Keble Road, Oxford, OX1 3RH, U.K. } \maketitle

\begin{abstract}
We use two independent methods to forecast the dark energy
measurements achievable by combining future galaxy redshift surveys
based on the radio HI emission line with Cosmic Microwave Background
(CMB) data from the {\sl Planck} satellite.  In the first method we
focus on the `standard ruler' provided by the baryon acoustic
oscillation (BAO) length scale. In the second method we utilize
additional information encoded in the galaxy power spectrum including
galaxy bias from velocity-space distortions and the growth of cosmic
structure.  We find that a radio synthesis array with about 10 per
cent of the collecting area of the Square Kilometre Array (SKA),
equipped with a wide ($10-100 ~ {\rm deg}^2$) field-of-view, would
have the capacity to perform a $20{,}000 ~ {\rm deg}^2$ redshift
survey to a maximum redshift $z_{\rm max} \sim 0.8$ and thereby
produce dark energy measurements that are competitive with surveys
likely to be undertaken by optical telescopes around 2015. There would
then be powerful arguments for adding collecting area to such a
`Phase-1' SKA because of the square-law scaling of survey speed with
telescope sensitivity for HI surveys, compared to the linear scaling
for optical redshift surveys. The full SKA telescope should, by
performing a $20{,}000 ~ {\rm deg}^2$ HI redshift survey to $z_{\rm
  max} \sim 2$ around 2020, yield an accurate measurement of
cosmological parameters independent of CMB datasets.  Combining CMB
({\sl Planck}) and galaxy power spectrum (SKA) measurements will drive
errors in the dark energy equation-of-state parameter $w$ well below
the 1 per cent level.  The major systematic uncertainty in these
forecasts is the lack of direct information about the mass function of
high-redshift HI-emitting galaxies.  `Stacking experiments' with SKA
pathfinders will play an important role in resolving this uncertainty.
\end{abstract}

\section{Introduction}

Over the past decade several different cosmological probes have
collectively demonstrated that the Universe contains a large fraction
of its energy density in clustered dark matter and unclustered dark
energy.  The analysis of the angular distribution of the Cosmic
Microwave Background (CMB) anisotropies requires that the total energy
density is close to that predicted in a spatially-flat Universe.  This
data, given a strong prior on the Hubble constant
\citep{2007ApJS..170..377S} imposed by the Hubble Key Project
\citep{2001ApJ...553...47F}, yields evidence for unclustered dark
energy. In a Universe that is close to spatially flat, observations of
distant Type-Ia supernovae provide independent evidence for a dark
energy component with a negative pressure driving an accelerated
expansion \citep{2006astro.ph.11572R}.  Furthermore, the analysis of
the shape of the galaxy power spectrum at low redshift requires a
matter density $\Omega_{\rm m} \sim$ 0.25
\citep[e.g.][]{2007ApJ...657..645P}, and the corresponding measurement
of baryon acoustic oscillations (BAOs) at $z=0.35$ agrees with this
picture \citep{2007ApJ...657...51P}.  Other probes confirm this
paradigm: a positive cross-correlation of the distribution of the
photons of the CMB with the distribution of galaxies at low redshift
indicates, via the Integrated Sachs-Wolfe (ISW) effect, that the
growth of large-scale over-densities is consistent with the presence
of dark energy with a negative pressure. The ISW effect has been
detected at the $> 2.5$-$\sigma$ level by several groups
\citep[e.g][]{2005NewAR..49...75B,
  2006MNRAS.372L..23C,2006astro.ph.10911R,2006PhRvD..74f3520G}.

No model has yet been successful in fitting all of the above
cosmological data without using a dark energy component. This does not
rule out a modified gravity model
\citep[e.g.][]{2006PhRvL..96a1301S,2006PhRvL..96d1103M} or more
complex cosmological assumptions such as back-reaction from
inhomogeneities \citep{2004JCAP...02..003R}, although dark energy from
back-reaction is disfavoured by other authors
\citep{2006CQGra..23..235I}.  However, the addition of a dark energy
component is the model that is the best-supported phenomenologically
to date, even though we do not have a clear understanding of its origin
or its nature.

Although there is a
wealth of information indicating the existence of
a dark energy component in our Universe, the measurements of its
properties, and specifically its equation of state $w = p/(\rho c^2)$
(where $p$ and $\rho c^2$ are the pressure and energy density of dark
energy respectively), are currently poor.  The simplest explanation
for dark energy is Einstein's cosmological constant, parameterized by
$w=-1$, although the observed magnitude of this vacuum energy poses a
formidable difficulty for theory.  Current measurements of the
parameter $w$ are at the 10\% level \citep{2007ApJS..170..377S}.

The discovery that $w \ne -1$ would yield vital information on the
nature of dark energy. If $w$ is a function of time (i.e.\ redshift)
and space, it could potentially approximate to a constant close to
$-1$ at late times, depending on its nature and evolution with
redshift \citep{2006PhRvD..73h3001S}.  The dark energy would therefore
behave as a cosmological constant at the present day.  We need to
measure both this equation-of-state parameter $w$ and its evolution
with time in order to prove or disprove whether dark energy is due to
the presence of a cosmological constant.

There are several types of observation which could in principle
achieve a measurement of $w$ with per-cent level accuracy, including
weak gravitational lensing, supernovae, and cluster number counts.
Future radio (HI emission) redshift surveys might make significant
contributions to this quest by measuring the large-scale distribution,
and hence clustering power spectrum, of galaxies
\citep{2004NewAR..48.1063B,2004NewAR..48.1013R,2005MNRAS.360...27A,2006MNRAS.367..169Z,2008PhRvL.100i1303C}.
The characteristic length scale of `wiggles' in the matter power
spectrum due to BAOs provides a set of `standard rulers' at any epoch
(i.e.\ redshift) at which they can be detected. BAO tests are
especially promising for measuring dark energy as they are believed to
be relatively immune to systematic effects
\citep{2003ApJ...594..665B,2003ApJ...598..720S,2003PhRvD..68f3004H}.
BAO measurements have already been successfully extracted from the
Sloan Digital Sky Survey (SDSS) dataset
\citep{2005ApJ...633..560E,2007ApJ...657...51P} and, although the
relatively small volume of the Universe probed by SDSS only permits a
measurement of $\Omega_{\rm m}$ rather than an accurate determination
of $w$, the prospects look extremely promising.

Developing a strategy for performing redshift surveys in radio
wavebands is very different from planning counterparts in optical
light.  We discuss in this paper the advantages and disadvantages of a
radio survey compared to an optical survey. We also make a prediction
of how much information can be obtained by measuring the full galaxy
power spectrum, and compare this to an analysis retaining only the
clean signature of the BAOs.

In Sec.\ref{sec:rad_vs_op} we discuss some of the differences between
optical and radio surveys of large-scale structure. In
Sec.\ref{sec:survey} we detail how we simulate HI surveys, also
referring the reader to \citet{2005MNRAS.360...27A}.  In
Sec.\ref{sec:bao_fit} we present forecasts for measuring dark energy
from several sizes of radio surveys that should, alongside CMB data
from the {\sl Planck}
satellite\footnote{http://www.rssd.esa.int/Planck/}, be attainable
over the coming decade. In Sec.\ref{sec:bao_mcmc} we use Monte Carlo
Markov Chain (MCMC) techniques to estimate the quality of future
galaxy power spectra; see also \cite{2007MNRAS.381.1313A}.  In
Sec.\ref{sec:chal} we comment on the mitigation of potential
systematic errors in the analysis.  We conclude our findings in
Sec.\ref{sec:conc}.

Unless otherwise stated we adopt a $\Lambda$CDM model with an extra
parameter $w$, a constant equation-of-state of dark energy.  Based on
recent measurements of cosmological parameters
\citep{2008arXiv0803.0547K,2008arXiv0803.0586D} we adopt the following
choice of parameters for our fiducial cosmological model:
\{$\Omega_b$,$\Omega_c$,$w$,$h$,$n_{\rm s}$,$\sigma_8$,$\tau$\} =
\{0.04,0.26,-1.0,0.72,1.0,0.9,0.09\}, where $\Omega_b$ and $\Omega_c$
are the present-day density of baryons and cold dark matter expressed
as a fraction of the critical density, $h = H_0/(100$ km s$^{-1}$
Mpc$^{-1}$) represents the Hubble Constant, $n_{\rm s}$ is the
primordial spectral slope of the power spectrum that has an overall
present-day normalization $\sigma_8$ and $\tau$ is the optical depth
to the last-scattering surface at recombination. We consider only
models that are spatially flat (i.e.\ $\Omega_{\Lambda}=0.7$ in our
fiducial cosmology) and we ignore the effects of neutrinos and other
forms of hot dark matter \citep{2007MNRAS.381.1313A}.

\section{A critical comparison of optical and radio redshift surveys}
\label{sec:rad_vs_op}

Undertaking `all-sky' galaxy redshift surveys in optical and radio
wavebands gives rise to very different issues.  Although astronomers
will be striving to complete both types of survey over the coming
decade, there is a natural competitive tension between the techniques
that is discussed in this Section.

The optical spectra of galaxies are much wealthier in information than
their radio spectra. Optical galaxy spectra exhibit absorption
features, and many have strong emission lines such as Ly$\alpha$,
[OII] and H$\alpha$, making redshift determination straightforward
with sufficient signal-to-noise.  The information is sufficiently rich
that one can pre-select a particular redshift range or type of galaxy
from the colours derived from a photometric imaging survey.  The
rest-frame equivalent widths of some emission lines can be very large
for subsets of galaxies, e.g. the star-forming galaxies, so blind
surveys are also plausible via `slit-less' techniques. Care must be
taken in both targeted and blind surveys to account for interlopers
from emission-line mis-identifications.

In the radio part of the spectrum the information in a galaxy spectrum
is much more limited.  The radio continuum is relatively featureless
allowing for only the very crudest photometric redshift techniques,
e.g.\ by using correlations between redshift and radio spectral index
\citep{2007MNRAS.375.1349C}. There is only one strong line in the
radio spectrum -- the `21-cm line' arising from hyperfine splitting in
the hydrogen atom (HI) -- and it is trivial for radio telescopes to
obtain the spectral dimension as part of an imaging survey.
Recent technological breakthroughs allow this to be achieved over
large frequency bandwidths. In principle, clean and blind redshift
surveys are therefore straightforward, with the luminosity of the HI
line providing a direct measure of the HI mass in an object.  HI line
surveys are hence similar to blind optical emission-line searches
performed with Integral Field Units (IFUs) or slit-less spectroscopy.

Unfortunately the HI line is extraordinarily weak.  Thus, at the time
of writing, all of the most important surveys of large-scale structure
have been undertaken with optical telescopes.  The state-of-the-art
method of performing these surveys is to use wide-field ($\sim \rm
deg^2$) multi-object spectrograph that, because of the multiplex
advantage of obtaining spectra of $10^{2-3}$ objects per pointing,
have been able to build up catalogues containing $10^{5-6}$ redshifts
using $10^{3-4}$ separate telescope pointings. With spectroscopic
exposures of the order of hours these programmes take $10^{2-3}$
clear dark observing nights and therefore many years to complete with
a single telescope. Surveys such as the SDSS have now mapped out
significant fractions ($\sim 1/4$) of the sky, but only at low
redshifts ($z \ltsimeq 0.5$) where moderate-sized (4-m class)
telescopes can measure absorption-line galaxy redshifts with relative
ease.

There are good prospects of pushing such studies to moderately-higher
redshifts ($z \ltsimeq 1$) with 4-m-class ground-based telescopes.
One technique is to target highly-biased tracers such as Luminous Red
Galaxies, which require a lower target density to suppress shot noise
in the power spectrum.  In this manner the Baryon Oscillation
Spectroscopic Survey (BOSS)\footnote{http://cosmology.lbl.gov/BOSS/},
an extension of the SDSS, will reach $z \sim 0.7$ over 25\% of the sky
by 2014. Another technique is to focus on objects that have strong
emission lines such as [OII], so that redshifts can be measured to a
greater depth without requiring a detection of the galaxy continuum
light.  In this manner the WiggleZ survey \citep{2007ASPC..379...72G}
at the Anglo-Australian Telescope (AAT) is probing galaxies up to $z
\sim 1$ over 2.5\% of the sky by 2010.  Both of these approaches
require a method of isolating specific types of high-redshift
galaxies, such as by using colour cuts and/or photometric redshifts.

Optical surveys can be performed to still higher redshifts using the
greater light grasp of 8-m class ground-based optical telescopes and
optimized observational strategies.  The recent deployment on the 8-m
Subaru Telescope of the FMOS near-IR multi-object spectrograph
\citep{2006SPIE.6269E.136D} will enable large-scale projects at
near-IR wavelengths, allowing redshift surveys based on
absorption-line (i.e. 4000-\AA\, break) systems at $z \gtsimeq 1$ and
emission-line (i.e.\ H$\alpha$) redshifts up to $z \sim 2$. There are
also plans for wide-field red-sensitive optical multi-object
spectrographs such as the proposed Wide Field Multi-Object
Spectrograph (WFMOS); this instrument would possess $> 1000$ fibres
and have sufficient sensitivity to push wide-field optical redshift
surveys to $z \sim 4$.

In principle, instruments like FMOS and WFMOS could survey the entire
sky for a selection of galaxies.  However, such surveys would be very
slow for three main reasons: (i) pre-selection of targets is hard, as
existing imaging surveys are either small or too shallow to select
targets; (ii) the field-of-view ($FOV$) of these spectrographs is of
order one square degree and exposures could take most of a clear dark
night at $z \gtsimeq 1.5$ (with the exception of line emitting
galaxies); and (iii) surveys must share time with other astronomical
programmes or instruments and are relatively limited in duration.  In
practice these future ground-based optical redshift surveys will
probably be limited to $\ltsimeq 2000 ~ \rm deg^2$ of sky. Similar
limitations on sky area apply to the alternative ground-based optical
technique of obtaining `blind' redshifts of high-redshift Ly$\alpha$
emitters using multiple IFUs deployed in the focal plane, such as
proposed by the HETDEX experiment \citep{2008arXiv0806.0183H}.

The most plausible approach for performing an all-sky high-redshift
spectroscopic galaxy survey in optical or infra-red wavebands is to
use a space-based platform, as suggested by several candidate
proposals for the Joint 
Dark Energy Mission (JDEM)\footnote{http://jdem.gsfc.nasa.gov/} funded in the
United States or the `Euclid' mission 
\citep{2008arXiv0807.4036R,2008ExA...tmp...37C} funded in Europe.  The extremely
low sky background from space enables much faster surveys, and galaxy
pre-selection can be avoided using slit-less techniques detecting
high-redshift H$\alpha$-emitting galaxies up to $z \sim 2$.  JDEM
could be launched by 2015 if funding continues.

Future radio telescopes provide an alternative route to constructing
all-sky galaxy redshift surveys up to $z \sim 2$.  Radio telescopes
benefit from a rapid rate of increase of survey speed with collecting
area.  The characteristic observing time $t_{\rm map}$ needed to
detect all the HI line-emitting galaxies above a fixed line-flux limit
over a fixed sky area scales with sensitivity and $FOV$ in accordance
with
\begin{equation}
\frac{1}{t_{\rm map}} \propto f^2 FOV
\label{eq:sr}
\end{equation}
where $f$ is the fractional sensitivity of the telescope, which we
define relative to a full SKA ($f \propto A_{\rm eff} / T_{\rm sys}$,
where $A_{\rm eff}$ is the effective collecting area and $T_{\rm sys}$
is the system temperature).  The quadratic nature of Eq.\ref{eq:sr}
implies a fundamentally different scaling to the corresponding {\sc
  etendue} figure used to characterize the survey speed of optical
telescopes ($1 / t_{\rm map} \propto A_{\rm eff} FOV$ in the case of
optical redshift surveys).  Future radio interferometers will possess
collecting areas which are orders of magnitude higher than those in
operation today, and the technology exists to produce a field-of-view
$FOV \sim 100$ deg$^2$, significantly larger than achievable by
optical spectrographs, and limited only by computational power.  The
large frequency bandwidth of the radio telescope will simultaneously
map out a significant redshift range; the maximum redshift attained is
determined by the sensitivity of the radio telescope and the degree of
concentration of interferometric baselines in the telescope core.

In summary, although radio wavebands offer very limited possibilities
for filtering techniques to pinpoint distant galaxies similar to those
commonly used in optical wavebands, the sheer survey volume that
future radio interferometers can map suggests that radio telescopes
have the potential to become very competitive facilities for
producing cosmologically-interesting redshift surveys.

A significant uncertainty in modelling future radio surveys is that
radio techniques have so far only yielded direct detections of HI
emission to $z \sim 0.2$
\citep{2001Sci...293.1800Z,2007arXiv0708.3853V,2008ApJ...685L..13C}.  
Various simple
models extrapolating the properties of HI galaxies to higher redshifts
are considered by \citet{2005MNRAS.360...27A}; we emphasize here the
critical role which SKA pathfinder telescopes will play in reducing
these uncertainties to the level needed to optimize the design of the
final SKA.  Pathfinders such as
FAST \citep{2008MNRAS.383..150D},
ASKAP\footnote{http://askap.pbwiki.com/} and
MeerKAT\footnote{http://www.ska.ac.za/} will achieve significant
expansions in frequency range and field-of-view over existing systems
with similar sensitivity (i.e. $A_{\rm eff} / T_{\rm sys}$).  One can
then envisage a set of `stacking experiments' that can yield
statistical measures of HI at much higher redshifts than the $z \sim
0.2$ objects currently under study. The $\sim 30 ~ \rm deg^2$ $FOV$
expected to be delivered by ASKAP will allow deep-field observations
covering (in $\sim 3$ pointings) $\sim 10^4$ `WiggleZ' star-forming
galaxies up to $z \sim 1$. Stacking the values of radio flux density
in channels corresponding to HI at the optical redshifts of these
objects provides, in the absence of any systematic noise floor, an
$\sqrt{10^4} \sim 100$-fold increase in effective sensitivity that
more than compensates for the factor $\sim 10$ diminution of line flux
expected between similar HI-emitting galaxies at $z \sim 0.2$ and $z
\sim 1$. The sensitivity to be delivered by MeerKAT over smaller
($\sim 1 ~ \rm deg^2$) sky patches will probe the general population
of $z \sim 1$ galaxies: FMOS Guaranteed Time Observations
\citep{2006SPIE.6269E.136D} expect to measure, via stellar absorption
features as well as H$\alpha$ emission, redshifts for $\sim 10^4$
galaxies per square degree.  These can also be stacked to achieve
high-redshift statistical detections of HI, provided that the noise in
the radio datasets is suitably well-behaved.

\section{Survey modelling}
\label{sec:survey}

We review here our method of simulating the HI surveys assumed in
Sec.\ref{sec:bao_fit} and Sec.\ref{sec:comb}.  In large part we have
followed the prescription of \citet{2005MNRAS.360...27A}, which we now
clarify in more detail (see also \citet{2006MNRAS.367..169Z}).

\subsection{HI mass limit}
\label{secHImass}

The r.m.s.\ sensitivity for a dual-polarization radio receiver at
system temperature $T_{\rm sys}$ for an integration of duration $t$ on a
telescope of effective collecting area $A_{\rm eff}$ is given by the
usual radio telescope equation\footnote{We note that the numerator of
  Eq.2 of \citet{2005MNRAS.360...27A} should read 100 mJy not 100
  nJy.}:
\begin{equation}
S_{\rm rms} = \frac{\sqrt{2} \, C \, k_B \, T_{\rm sys}}{A_{\rm eff}
  \, \sqrt{\Delta \nu \, t}} \sim \left( \frac{1.6 \, \mu{\rm Jy}}{f}
\right) \sqrt{ \frac{1}{\Delta \nu / {\rm MHz}} } \sqrt{ \frac{1}{t /
    {\rm hr}} }
\label{eqsrms}
\end{equation}
where $C$ is a constant of order unity \citep[see Chapter 1 of
][]{2001isra.book.....T}, $k_B$ is Boltzmann's constant, $\Delta \nu$
is the frequency bandwidth, and $f = [A_{\rm eff} /(10^6 \, {\rm
    m}^2)] [T_{\rm sys} / 50 ~\rm K]^{-1}$.  The value of $f$ hence
parameterizes the sensitivity of the radio telescope compared to the
SKA.

In practice we used for our calculations the frequency-dependent
telescopes sensitivity listed in Table 1.2 of
\citet{1999sska.conf.....T} which we express in the following fitting
formula:
\begin{eqnarray}
S_{\rm rms} &=& \frac{(1.790365 \, \nu^2 - 4.53125 \, \nu + 4.46666) \,
  \mu{\rm Jy}}{f} \nonumber \\ &\times& \sqrt{ \frac{30 \, {\rm
      km/s}}{\Delta V} } \sqrt{ \frac{4 \, {\rm hr}}{t} }
\label{eqsrms2}
\end{eqnarray}
where $\nu$ is the frequency of the redshifted HI line.  The first
term in Eq.\ref{eqsrms2} is equivalent to a dual-polarization
r.m.s.\ sensitivity in a velocity channel of width $\Delta V = 30$ km
s$^{-1}$ in a $t = 4$ hr exposure.  The underlying assumptions are
that the scaling of $S_{\rm rms}$ with telescope sensitivity $f$ is
independent of frequency, and $C=\sqrt{2}$, which makes the optimistic
assumption that all statistically-independent signal samples can be
successfully averaged over a given exposure time.  The assumption that
$f$ is independent of frequency implies that the increasing
contribution of sky temperature to $T_{\rm sys}$ at lower frequencies
is compensated for by an increase in $A_{\rm eff}$ with decreasing
frequency.  Given that the brightness temperature of the Galactic
foreground emission scales roughly as $T_{\rm Gal} \sim 20 (\nu /
408~MHz) ^{-2.7} ~K$, in practice this will become increasingly
difficult to realize below $\sim 300 ~ \rm MHz$ (HI at $z \sim 3.7$)
once the rapidly rising sky temperature begins to dominate the system
temperature.

The flux detection limit $S_{\rm lim}$ for galaxies is defined by the
threshold parameter
\begin{equation}
n_\sigma = S_{\rm lim}/S_{\rm rms}.
\label{eqsn}
\end{equation}

We allow for a certain fraction of flux to be `resolved out' using a
model for the source sizes and the interferometric baseline
distribution.  The proper physical size of the sources is modelled as
$R(z) = R_0/(1+z)^2$ following the discussion in AR05, where $R_0 =
15$ kpc.  The fraction of flux detected from a source at redshift $z$
(see also Fig.2 of AR05) is then described as an integral over the
distribution $f(B)$ of baselines $B$:
\begin{equation}
{\rm Frac \; detected} = \int_{0 \, {\rm km}}^{10000 \, {\rm km}} f(B)
\exp{ \left( - \frac{B^2}{2B_0(z)^2} \right) } \, {\rm d}B
\end{equation}
where $B_0 = [\lambda x(z)]/[\pi R(z)]$ and $\lambda = 0.21$ m, $x(z)$
is the co-moving radial distance to redshift $z$, and the baseline
distribution $f(B)$ is given by
\begin{equation}
f(B) = \sum_i a_i \sqrt{\frac{2}{\pi}} \frac{1}{B_i} \exp{ \left( -
  \frac{B^2}{2B_i^2} \right) }
\end{equation}
where we use a sum of four different baseline components $i=1$ to
$i=4$ where $B_1 = 0.5$ km, $B_2 = 3$ km, $B_3 = 150$ km and $B_4 =
2600$ km (this latter corresponding to $0.02$ arcsec resolution for
wavelength $\lambda = 0.21$ m).  The coefficients $a_i$ satisfy
$\sum_{i=1}^{i=4} a_i = 1$ (such that $\int f(B) \, {\rm d}B = 1$) and
were taken as $a_1 = 0.2242$, $a_2 = 0.2746$, $a_3 = 0.2506$, $a_4 =
0.2506$. This is a composite array and each $a_i$ component of the
array is distributed in a `scale-free' configuration, as used in AR05.
We note that the costs of data transport increase dramatically for the
longest baselines of a radio interferometer and thus the final
realization of the SKA is likely to be more centrally concentrated
than we have assumed, meaning that we are being conservative in our
assumed ability of the SKA to detect HI in galaxies.

The width of the HI emission line determines the relevant frequency
bandwidth $\Delta \nu$ in Eq.\ref{eqsrms} and is related to a
line-of-sight velocity width $V(z)$ at redshift $z$:
\begin{equation}
\Delta \nu = \left( \frac{\nu_0}{1+z} \right) \frac{V(z)}{c},
\label{eqwidth}
\end{equation}
where $\nu_0 = 1.4204$ GHz is the rest-frame frequency of the HI line,
$c$ is the speed of light and $V(z)$ is measured in the rest-frame of
the galaxy.  We follow AR05 in assuming a scaling with redshift for
the typical rest-frame velocity width of the line $V(z) =
{V_0}/{\sqrt{1+z}}$, where $V_0 = 300$ km s$^{-1}$, and ignore the
effects of inclination.

The luminosity of HI photons from a gas cloud of mass $M_{\rm HI}$ can
be derived using atomic physics (see, e.g., Sec.2.2 of AR05).
Assuming a flat profile for the emission line with frequency, we
obtain the following equation linking $M_{\rm HI}$ with the observed
flux density $S_\nu$:
\begin{equation}
M_{\rm HI} = \frac{16 \pi}{3} \frac{m_H}{A_{12} hc} \, x(z)^2 \, S_\nu \,
V(z) \, (1+z) 
\end{equation}

\noindent or in usual units:

\begin{equation}
M_{\rm HI} = 0.235 {M_\odot} \, (x / {\rm Mpc})^2 \,
(S_\nu / \mu{\rm Jy}) \, (V / {\rm km s^{-1}}) \, (1+z)
\end{equation}

\noindent where $m_H$ is the atomic mass of hydrogen, $A_{12}$ is the Einstein
coefficient for the transition, $h$ is Planck's constant.  and $x(z)$
is the co-moving distance to redshift $z$.

\subsection{HI mass function modelling}
\label{sec:HImass}

\begin{figure}
\center
\includegraphics[width=8.5cm,angle=0]{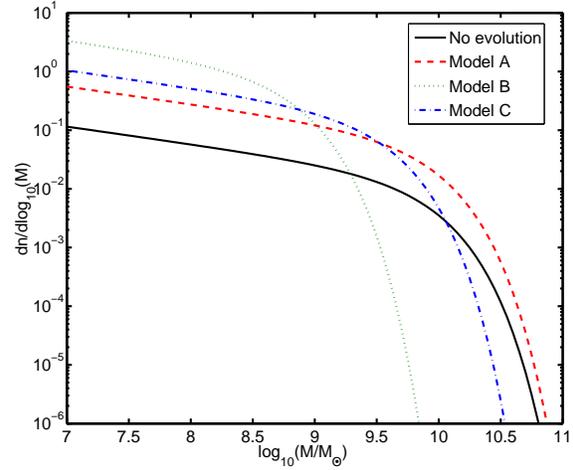}
\caption{ The HI mass function plotted at redshift $z=2$ for a
  no-evolution model (black solid line), evolutionary models A (red
  dashed line), B (green dotted line) and C (blue dot-dashed
  line). This is a new version of Fig.4 from AR05 which was incorrect
  due to missing factors of $h$ in the plotting routine. The
  evolutionary models are described in Sec.~\ref{sec:HImass}.  }
\label{figHImass}
\end{figure}

We assume that the HI mass function at all redshifts is described by a
Schechter function for the number density of galaxies ${\rm d}n$ in a
mass range ${\rm d}M$:
\begin{equation}
\frac{{\rm d}n}{{\rm d}M} \, {\rm d}M = 
\phi^* \left( \frac{M}{M^*} \right)^\alpha \exp{
  \left[ - \left( \frac{M}{M^*} \right) \right] } \, \frac{{\rm d}M}{M^*}
\end{equation}
parameterized by a low-mass slope $\alpha$, a characteristic mass
$M^*$ and a normalization $\phi^*$.  This function has only been
measured in the local Universe.  The HIPASS survey reported the
results: $\alpha = -1.3$, $M^* = 3.47 \, h^{-2} 10^9 \, M_\odot$,
$\phi^* = 0.0204 \, h^3$ Mpc$^{-3}$
\citep{2003AJ....125.2842Z}\footnote{We note that due to missing
  factors of $h$ in a plotting routine, Fig.4 of
  \citet{2005MNRAS.360...27A} is incorrect, and it should be replaced
  by Fig.\ref{figHImass} of this paper.}.  Throughout this study we
assume $\alpha = -1.3$ at all redshifts.  Our survey mass limit
typically lies above $M^*$, thus our results are largely insensitive
to the value of $\alpha$.  The HI mass function model must specify the
redshift-evolution of $M^*(z)$ and $\phi^*(z)$.

By integrating the mass function we can deduce the overall mass
density of neutral hydrogen at a given redshift:
\begin{equation}
\rho_{\rm HI}(z) = \phi^*(z) \, M^*(z) \, \Gamma(\alpha + 2),
\end{equation}
where $\Gamma$ denotes the Gamma function.  We used the HIPASS survey
mass function measurements to fix the value of $\rho_{\rm HI}$(0).  In
order to model the evolution of $\rho_{\rm HI}$ with redshift, we
followed AR05 and employed the measurements of the neutral gas density
in damped-Ly$\alpha$ (DLA) systems as a function of redshift by
\citet{2001ApSSS.277..551P}. Specifically, we used the following
functional form produced by AR05 by fitting to DLA data points, making
a somewhat ad-hoc correction for selection effects in the
damped-Ly$\alpha$ studies:
\begin{equation}
\Omega_{\rm HI}(z) = N \left[ 1.813 - 1.473 \, (1+z)^{-2.31} \right]
\label{eqhinorm}
\end{equation}
where $\Omega_{\rm HI}(z) = \rho_{\rm HI}(z)/\rho_c(z=0)$ is the HI
mass density relative to the critical density of the Universe
($\rho_c(z=0) = 2.7755 \, h^2 10^{11} \, M_\odot$ Mpc$^{-3}$).  The
normalization constant is fixed by the zero-redshift HIPASS results.

\begin{figure}
\center
\includegraphics[width=8cm]{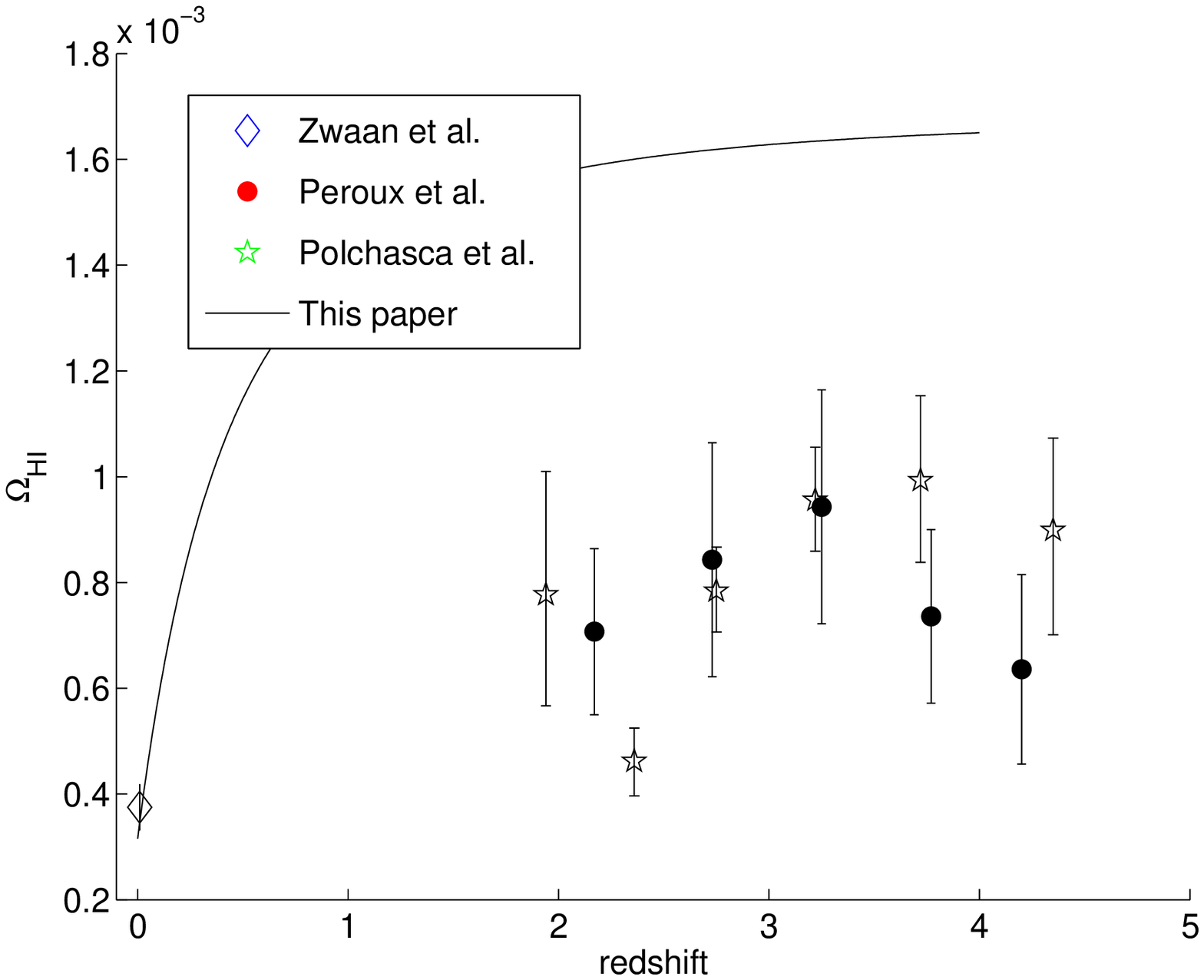}
\caption{Measurements of the evolution of the HI mass density relative
  to the critical density of the Universe, $\Omega_{\rm HI}(z)$,
  derived from HI emission studies \citep{2005MNRAS.359L..30Z} and
  studies of damped-Ly$\alpha$ absorption lines in distant quasars
  \citep{2005ApJ...635..123P,2005MNRAS.363..479P}.  We have corrected
  all data to our fiducial cosmological model.  The solid line shows
  the evolutionary behaviour assumed in this paper, we have assumed following
  AR. This line adopts the maximal
  correction (factor ~2) inferred from radio-selected samples
  for dust biasing against the optical detection of HI-rich systems in
  absorption \citep{2001A&A...379..393E}. 
  A recent study combining absorption-system
  results from optical- and radio-selected samples \citep{2008arXiv0810.3236P}
  suggests that the correction factor is unlikely to be much more than ~1.3
  subject to the caveat that a complete understanding of this factor requires
  more careful consideration of the effects of gravitational lensing, as well
  as dust, bias.}
\label{figsteve}
\end{figure}

If we use Eq.\ref{eqhinorm} to determine the redshift-variation of the
normalization of the mass function, then the full evolution model may
be specified by the behaviour of the break in the Schechter function,
$M^*(z)$.  AR05 consider four different possibilities that, subject
to the observational uncertainties in HI evolution detailed in the
caption to Fig.\ref{figsteve}, span the plausible range of behaviours.
We summarize these here, referring the reader to AR05 for a more
detailed discussion:

\begin{itemize}

\item {\bf No-evolution model:} $\phi^*$ and $M^*$ remain equal to the
  zero-redshift values measured by the HIPASS survey.

\item {\bf Model A:} $\phi^*$ scales with $z$ using the DLA results
(Eq.\ref{eqhinorm})
  with $M^*(z) = M^*(0) =$ assumed constant.

\item {\bf Model B:} $\phi^*$ scales with $z$ using the DLA results.
  $M^*(z) = M^*(0) \, D(z)^3$, where $D(z)$ is the linear growth
  factor at redshift $z$ normalized such that $D(0) = 1$.

\item {\bf Model C:} $\phi^*$ scales with $z$ using the DLA results.
  The break in the mass function is additionally controlled by the
  cosmic star formation rate.

\end{itemize}

In model C the break in the mass function evolves as
\begin{equation}
\frac{M^*(z)}{M^*(0)} = \left( \frac{ \left[ \Omega_{\rm
      stars}(0)/\Omega_{\rm HI}(0) \right] + 2}{ \left[ \Omega_{\rm
      stars}(z)/\Omega_{\rm HI}(z) \right] + 2} \right) D(z)^3
\label{eqmodelc}
\end{equation}
where the fractional density in stars $\Omega_{\rm stars}(z) =
\rho_{\rm stars}(z)/\rho_c(z=0)$ is deduced using the fits to the
cosmic star-formation history in \citet{2002ApJ...574...59C} and there
is an implicit assumption that $\Omega_{\rm HI}(z) = \Omega_{\rm
  H_2}(z)$.  Specifically,
\begin{equation}
\Omega_{\rm stars} (z) = \frac{1}{\rho_c(z=0)} \int_z^\infty \left[
  \frac{SFR(z^{\prime})}{H(z^{\prime}) \, (1+z^{\prime})} \right] \,
      {\rm d}z^{\prime}
\end{equation}
where $H(z)$ is the Hubble parameter at redshift $z$ and we take
\begin{equation}
SFR(z) = \left[ \frac{0.13}{1 + 6 \exp{(-2.2 z)}} \right] \left[
  \frac{2}{2.5} \right] \left[ \frac{x_{\rm fid}(z)}{x(z)} \right]
\label{eqsfr}
\end{equation}
where $x(z)$ is the co-moving distance to redshift $z$ and $x_{\rm
  fid}(z)$ is the value of $x(z)$ assuming a fiducial cosmology with
$\Omega_{\rm m} = 1$ to convert from the cosmology assumed by
\citet{2002ApJ...574...59C}.  The units of $SFR(z)$ in Eq.\ref{eqsfr}
are $h^2 M_\odot$ yr$^{-1}$ Mpc$^{-3}$. There
is an implicit assumption that $\Omega_{\rm HI}(z) = \Omega_{\rm H_2}(z)$; 
although $\Omega_{\rm HI} \sim 4\Omega_{H_2}$ is more appropriate
at low redshifts (Obreshkow \& Rawlings 2008), model predictions of the cosmic
evolution in the degree of molecularization in galaxies (Obreshkow et al., 
in prep) suggest that a rough equality becomes a reasonable assumption at 
$z \gtsimeq 1$. Our default assumption will
be evolution Model C, following the recommendation of AR05.

\subsection{Survey parameters}

We consider HI emission-line surveys defined by four parameters:

\begin{itemize}

\item The telescope sensitivity as a fraction of that achieved by the
  SKA (the parameter $f$ in Eq.\ref{eqsrms}).

\item The survey detection threshold $n_\sigma$ in Eq.\ref{eqsn}.

\item The telescope field-of-view at the HI emission-line frequency at
  zero redshift, $FOV$.

\item The assumed HI evolution model.

\end{itemize}

The scaling of the observed wavelength of the HI emission line with
redshift can produce a corresponding scaling of the field-of-view
$FOV$ of the radio telescope, depending on the telescope design.  We
assume for now parabolic dishes with single-pixel feeds, 
such that $FOV \propto (1+z)^2$.  This
scaling may be used in a survey tiling scheme to produce an effective
increase with redshift in the integration time $t$ of the HI survey,
and this is incorporated into our model following AR05.

We assume an overall survey duration equal to 1 yr covering
approximately one hemisphere, $20{,}000$ deg$^2$.  We do not consider
trade-offs between total survey area and redshift depth in this paper.

The bandwidth of the radio telescope may not be sufficient to cover
the wavelength range of the HI emission line up to the maximum
observable redshift $z_{\rm max}$, necessitating repeat observations
of each sky area.  We use a parameter $\beta$ to represent the ratio
of the telescope bandwidth to the required frequency range of HI
observations if we assume constant exposure time in each frequency
block (in practice more distant redshift slices will require longer
integrations and an `effective' value of $\beta$ will result).  For
example if $\beta \times FOV = 10$ deg$^2$ then a $20{,}000$ deg$^2$
survey requires 2000 pointings; if $\beta = 1$ each is of 4-hours duration, 
and this
requires 1-year exposure time (adopting a 10 per cent overhead for
calibration and other operational tasks). We note that, for a given
survey duration and total solid angle, Eq.\ref{eqsrms} shows that the
survey is identical if the combination $(\beta \, FOV \, f^2 /
n_\sigma)$ remains unchanged.  For example, halving the detection
threshold $n_\sigma$ is equivalent to doubling the field-of-view $FOV$
or multiplying $f$ by $\sqrt{2}$.

In subsequent calculations we adopt $f=0.15$ as indicative of the
sensitivity of an instrument representing `phase-I' of the SKA (which
may be available in 2015). Although it is anticipated that such an
instrument will have $\sim 10$ per cent of the final collecting area
of the SKA, it is likely for practical reasons to have a much larger
fraction of its baselines within the compact core than is predicted by
the baseline distribution function of Sec.\ref{secHImass}.  Hence the
sensitivity will be higher for fixed $f$ as a greater fraction of
baselines are useful for source detection.

\subsection{Clustering of HI-emitting galaxies}
\label{secclust}

We assume that the HI-emission galaxy power spectrum on large scales
$P_{\rm HI}(k)$ is related to the underlying linear dark matter power
spectrum $P_{\rm lin}(k)$ by a constant bias factor $b$, as is
observed for all classes of galaxy on large scales:
\begin{equation}
P_{\rm HI}(k) = b^2 \, P_{\rm lin}(k)
\end{equation}

HI-emitting galaxies are known to be one of the least biased tracers
of mass over-densities (i.e.\ to possess one of the lowest values of
$b$) owing to their avoidance of dense regions such as galaxy
clusters.  The clustering of the zero-redshift HIPASS survey has been
measured by \citet{2007MNRAS.378..301B} by fitting a two-point
correlation function parameterized by a power-law $\xi(r) =
(r/r_0)^{-\gamma}$, with clustering length $r_0$ and slope $\gamma$.
The results were $r_0 = 4.1 \pm 0.3 \, h^{-1}$ Mpc (in redshift space)
and $r_0 = 3.1 \pm 0.3 \, h^{-1}$ Mpc (in real space), using the fixed
canonical value $\gamma = 1.8$.  This corresponds to a linear bias
parameter $b = 0.63$.

No equivalent measurement exists at higher redshift, but we note that
many classes of galaxy evolve in their clustering such that $r_0(z)
\approx$ constant where $r_0$ is measured in co-moving co-ordinates
\citep[e.g.][]{2002MNRAS.333..961L}.  This forces a corresponding evolution in
the linear bias $b \propto 1/D(z)$. Bias appears, however, to be a
strong function of HI mass, in fact \citet{2007MNRAS.378..301B} suggest
that the more massive galaxies have a bias similar to optical galaxies,
i.e. $b=1$.  For the purposes of this study and
because of the lack of high-redshift measurements, we assume $b=1$
independent of redshift; this is consistent with the clustering
strength of objects near the break in the $z=0$ HI mass function.

The clustering amplitude is significant in fixing the relative
contribution of cosmic variance and shot noise to the error in a
measurement of the power spectrum.  These two sources of error are
equal (at a given Fourier scale $k$) when the quantity $n \times
P_{\rm HI}(k) = 1$, where $n$ is the galaxy comoving number density.

\subsection{Redshift distributions}

\begin{figure}
\center
\includegraphics[width=6cm,angle=-90]{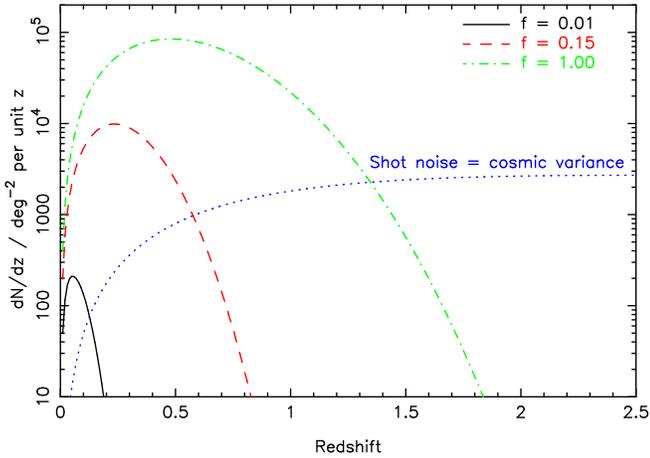}
\caption{The dependence of the HI galaxy redshift distribution ${\rm
    d}N/{\rm d}z$ on the fractional sensitivity $f$ of the radio
  telescope (where $f=1$ for the SKA), assuming 1 yr of observing and
  $\beta \times FOV = 10$ deg$^2$.  Other important assumptions are:
  (i) $FOV$ scales with redshift as $(1+z)^2$; (ii) the baseline
  distribution and galaxy size assumptions of Sec.\ref{secHImass} are
  adopted; (iii) HI evolution follows AR05 model C; (iv) only galaxies
  detected with a significance exceeding 10$\sigma$ are considered.}
\label{figdndzarea}
\end{figure}

\begin{figure}
\center
\includegraphics[width=6cm,angle=-90]{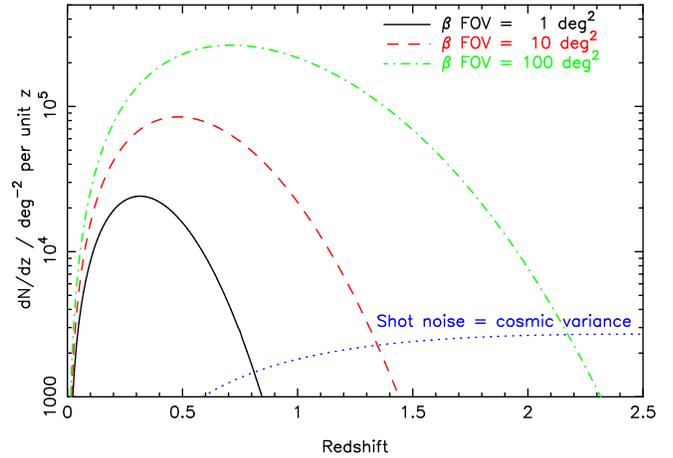}
\caption{The dependence of the HI galaxy redshift distribution ${\rm
    d}N/{\rm d}z$ on the quantity $\beta \times FOV$, assuming 1 yr of
  observing and a telescope sensitivity $f=1$.  See the caption to
  Fig.\ref{figdndzarea} for other assumptions.}
\label{figdndzfov}
\end{figure}

Figs.\ref{figdndzarea} and \ref{figdndzfov} plot the HI galaxy
redshift distribution $dN/dz$ for various different survey parameters
assuming detection threshold $n_\sigma = 10$.  The source density
corresponding to $n \, P_{\rm HI}(k) = 1$ (for our assumed HI
clustering properties) is also indicated, using the characteristic
scale $k = 0.15 \, h$ Mpc$^{-1}$. The redshift at which ${\rm d}N/{\rm
  d}z$ crosses this locus indicates the maximum redshift at which the
error in the clustering measurements is dominated by cosmic variance
rather than by shot noise.

In Fig.\ref{figdndzarea} we assume $\beta \, FOV$ = 10 deg$^2$,
$n_\sigma = 10$ and the preferred evolution model C, and plot ${\rm
  d}N/{\rm d}z$ for fractional SKA areas $f = 0.01$, $f = 0.15$ and $f
= 1$.  In Fig.\ref{figdndzfov} we fix $f = 1$ and plot ${\rm d}N/{\rm
  d}z$ for $\beta \, FOV = 1$, $10$ and $100$ deg$^2$.  Note that an
identical redshift distribution is produced if the survey parameters
are changed such that the combination $(\beta \, FOV \, f^2 /
n_\sigma)$ is left invariant.

\subsection{Effective survey volume}

A useful `figure-of-merit' for the accuracy with which a galaxy survey
spanning a range of redshifts can measure the power spectrum at a
given scale $k$ is the {\it effective volume}.  This is the cosmic
volume weighted by a redshift-dependent shot-noise factor:
\begin{equation}
V_{\rm eff}(k) = \int_0^\infty \left[ \frac{n(z) P_{\rm HI}(k,z)}{1 +
    n(z) P_{\rm HI}(k,z)} \right]^2 \frac{{\rm d}V}{{\rm d}z} \, {\rm d}z
\label{eqveff}
\end{equation}
In our assessment we take $k = 0.15 \, h$ Mpc$^{-1}$ as the relevant
scale, because this defines the approximate extent of the linear
regime at $z \approx 0.5$ and also characterizes the regime of
interest for the baryon oscillations.

\begin{figure*}
\center
\includegraphics[width=10.5cm,angle=-90]{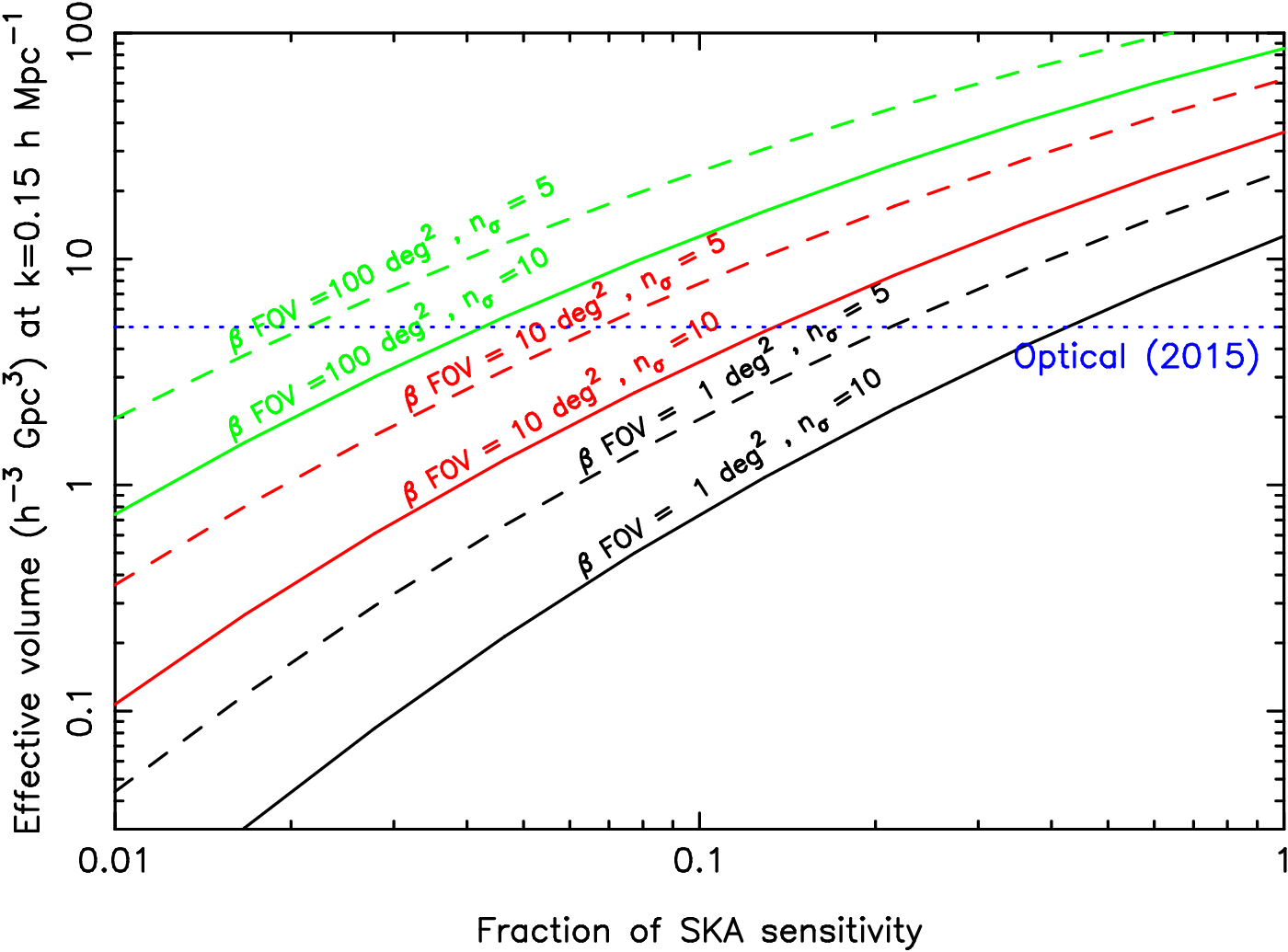}
\caption{The dependence of the effective HI survey volume achieved in
  1 yr of observing on the fraction of SKA sensitivity, for various
  different values of field-of-view ($\beta \times FOV$) and detection
  threshold $n_\sigma$. See the caption to Fig.\ref{figdndzarea} for
  other assumptions. A rough benchmark is plotted for future
  ground-based redshift surveys in optical and near-infrared
  wavebands.
\label{figveffsurv}}
\center
\includegraphics[width=10.5cm,angle=-90]{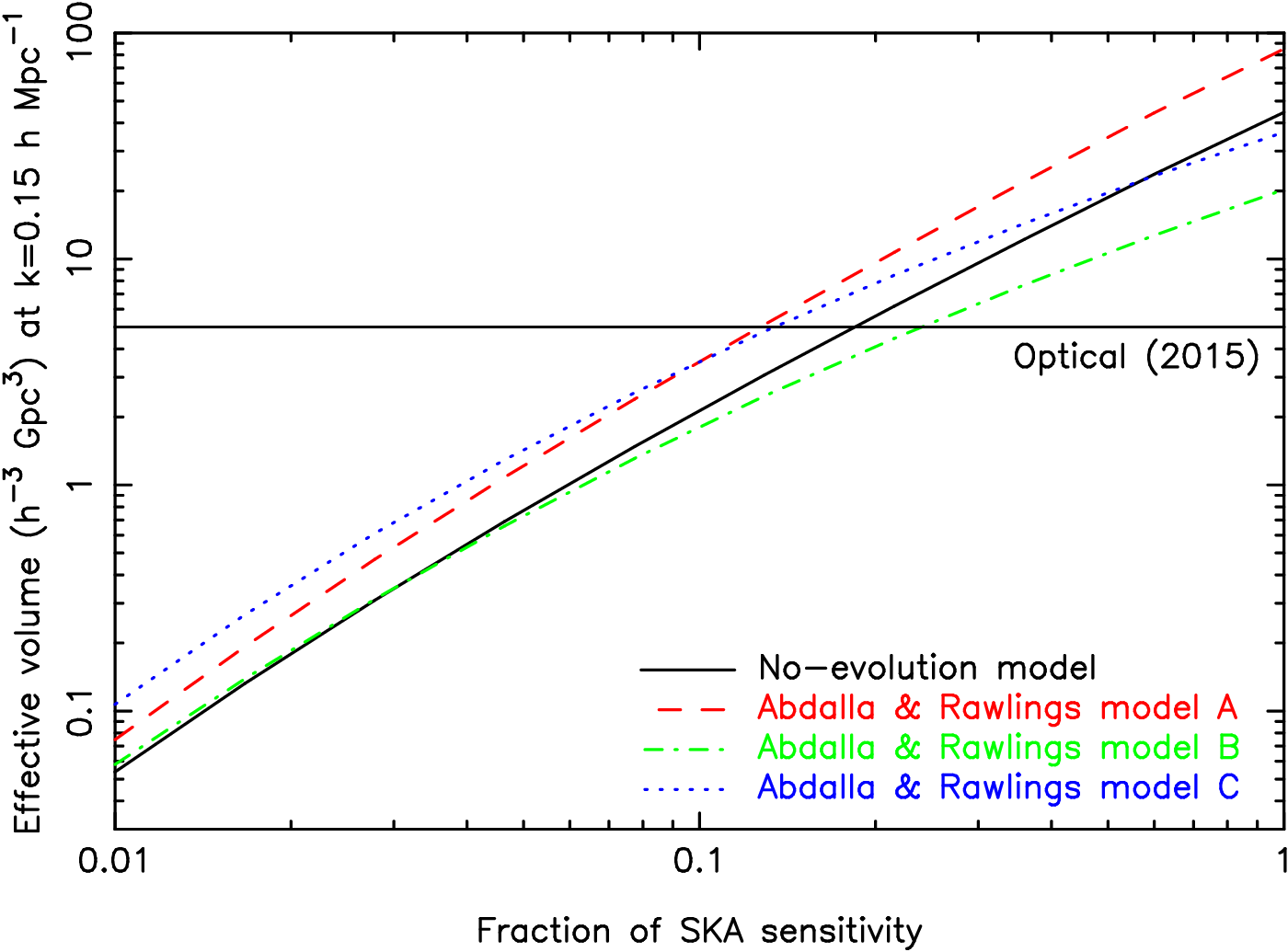}
\caption{The dependence of the effective HI survey volume achieved in
  1 yr of observing on the fraction of SKA sensitivity, for different
  HI evolution models. We assume $\beta \times FOV = 10$ deg$^2$; see
  the caption to Fig.\ref{figdndzarea} for other assumptions.  A rough
  benchmark is plotted for future ground-based redshift surveys in
  optical and near-infrared wavebands.
\label{figveffevol}}
\end{figure*}

In Figs.\ref{figveffsurv} and \ref{figveffevol} we display the
variation of the HI survey effective volume with the telescope
sensitivity $f$ over the range from $f = 0.01$ to $f = 1$.  For
comparison we also indicate the approximate volume that is likely to
be probed by ground-based surveys in optical wavebands on a timescale
of 2015.  We base this benchmark -- $V_{\rm eff} = 5 \, h^{-3}$
Gpc$^3$ -- on a projection of results from the Baryon Oscillation
Spectroscopic Survey (BOSS) or the proposed Wide Field Multi-Object
Spectrograph (WFMOS).  This volume is equivalent to a survey of
$10{,}000$ deg$^2$ covering $z < 0.7$, or 2000 deg$^2$ spanning $0.5 <
z < 1.5$ (if we neglect the shot noise factor in Eq.\ref{eqveff}).  We
note that assuming near-infra red spectroscopy, space-based surveys
such as Euclid or JDEM would cover galaxies a redshift range of roughly
$z < 2.0$ \citep{2008ExA...tmp...37C}, 
hence a similar volume achievable by a full SKA.

In Fig.\ref{figveffsurv} we assume evolution model C and consider six
different combinations of field-of-view and detection threshold.  We
note that the minimum sensitivity required by a radio telescope to
equal the effective volume of our future optical survey benchmark
(assuming $n_\sigma = 10$) are:
\begin{itemize}
\item $\beta \, FOV = 1 \, {\rm deg}^2 \rightarrow f = 0.4$
\item $\beta \, FOV = 10 \, {\rm deg}^2 \rightarrow f = 0.13$
\item $\beta \, FOV = 100 \, {\rm deg}^2 \rightarrow f = 0.04$
\end{itemize}
This is consistent with the quantity $FOV \times f^2$ characterizing
the survey capability of radio telescopes.

In Fig.\ref{figveffevol} we assume $\beta \, FOV = 10$ deg$^2$ and
$n_\sigma = 10$ and display the dependence of the effective volume on
the HI evolution model.  The above conclusions regarding required
collecting area are not strongly changed if a different evolution
model is considered.  In the most pessimistic case, an extra $\approx
50\%$ of sensitivity is required.

\subsection{Maximum survey redshift}
\label{seczmax}

\begin{figure*}
\center
\includegraphics[width=10.5cm,angle=-90]{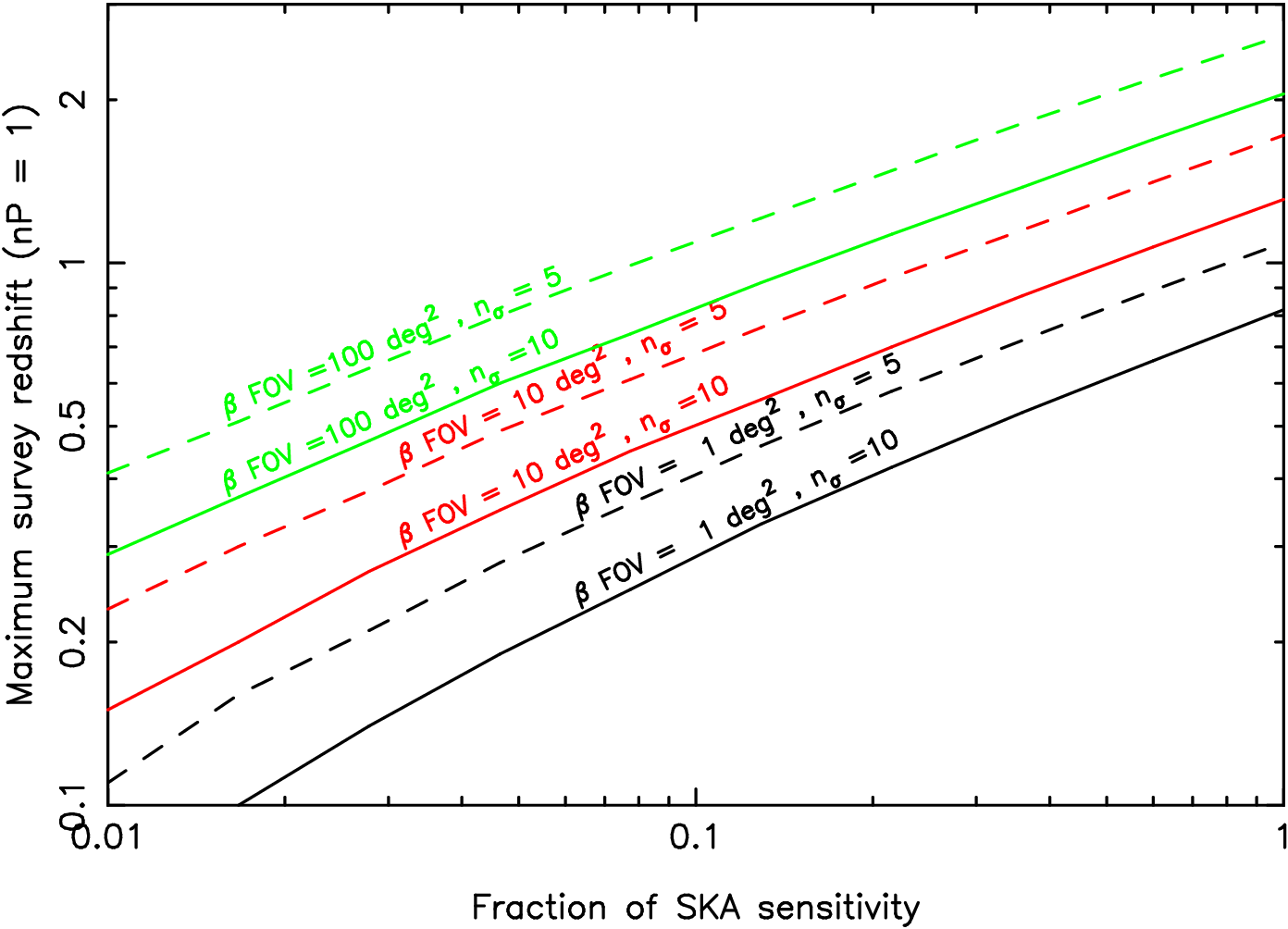}
\caption{The dependence of the maximum HI survey redshift $z_{\rm
    max}$ achieved in 1 yr of observing on the value of $\beta \times
  FOV$ and detection threshold $n_\sigma$. See the caption to
  Fig.\ref{figdndzarea} for other assumptions.
\label{figzmaxsurv}}
\center
\includegraphics[width=10.5cm,angle=-90]{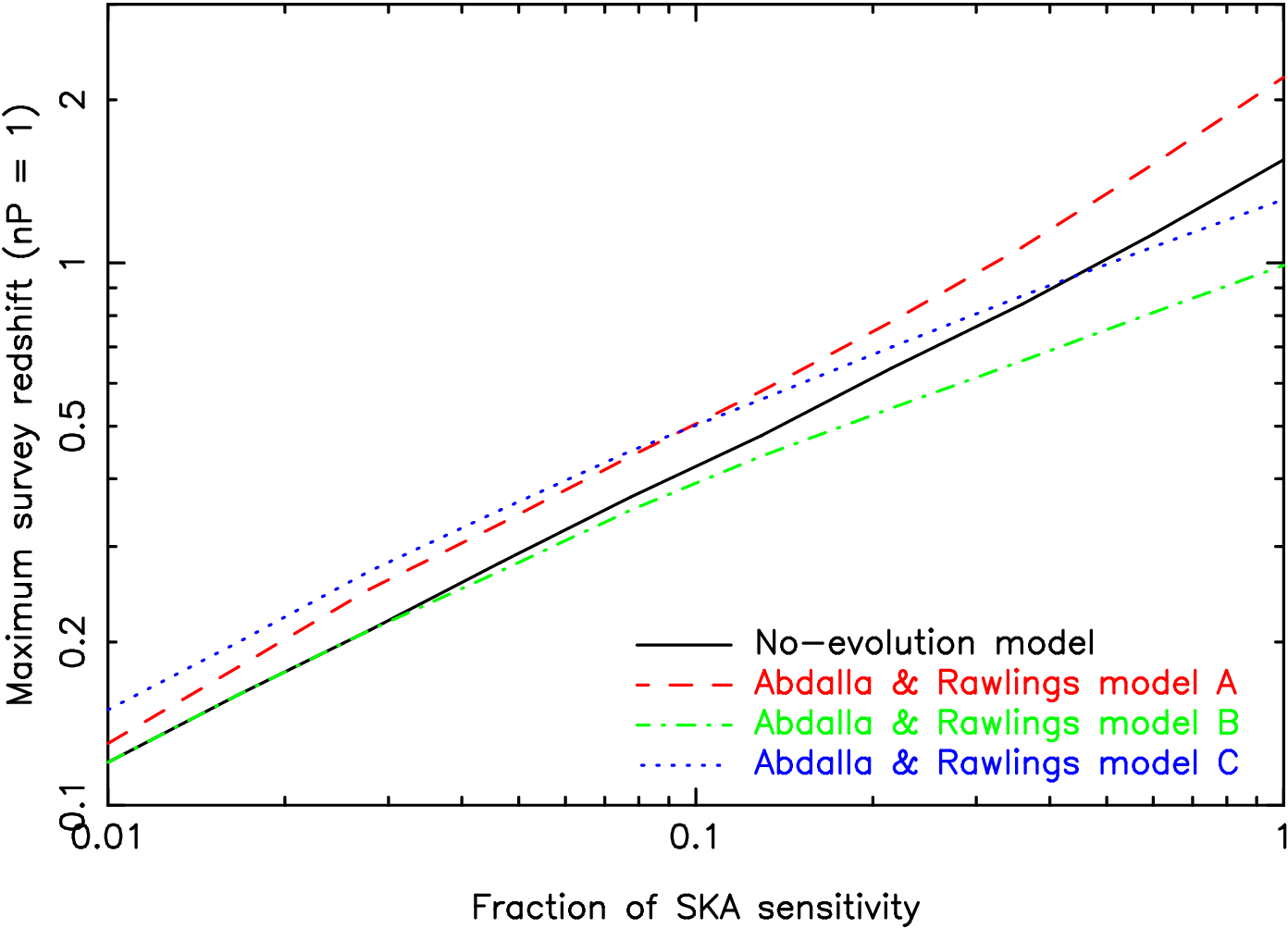}
\caption{The dependence of the maximum HI survey redshift $z_{\rm
    max}$ achieved in 1 yr of observing on the fraction of SKA
  sensitivity, for different HI evolution models.  A telescope with
  $\beta \times FOV = 10$ deg$^2$ is assumed; see the caption to
  Fig.\ref{figdndzarea} for other assumptions.
\label{figzmaxevol}}
\end{figure*}

We used the criterion $n \, P_{\rm HI}(k) = 1$ to estimate the maximum
redshift $z_{\rm max}$ reachable in our model survey.
Figs.\ref{figzmaxsurv} and \ref{figzmaxevol} display the dependence of
this maximum redshift on the telescope collecting area in the same
style as Figs.\ref{figveffsurv} and \ref{figveffevol}.  The HI radio
surveys which match the performance of our benchmark future
ground-based optical survey correspond to all-hemisphere surveys up to
a maximum redshift $z \approx 0.6$ for an $f = 0.15$ radio array like
the Phase-I SKA instrument. Surveys with a full SKA would access
redshifts significantly beyond $z = 1$, perhaps reaching $z \sim 2$.
Again, these conclusions do not depend strongly on the assumed HI
evolution model.

\section{Simulated dark energy measurements using baryon oscillation fitting}
\label{sec:bao_fit}

\subsection{Survey power spectra}

\begin{figure*}
\center
\includegraphics[width=10.5cm,angle=-90]{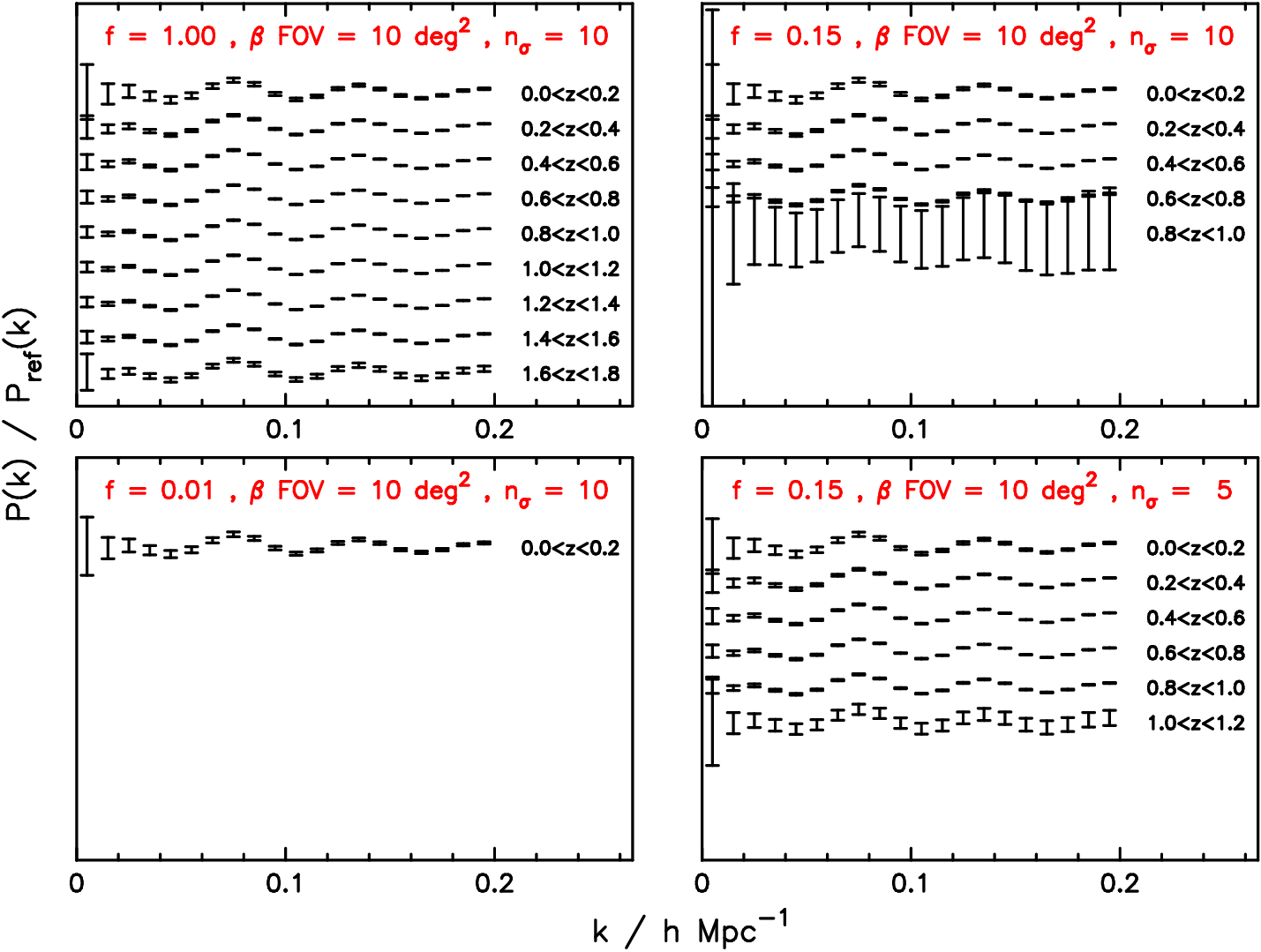}
\caption{Power spectrum simulations for a series of redshift slices
  for four different HI survey parameter sets.  The power spectra are
  divided by a smooth reference spectrum to emphasize the baryon
  acoustic oscillations.  Power spectra for different redshift slices
  are offset vertically for clarity.  Points are plotted up to a
  maximum value $k = 0.2 \, h$ Mpc$^{-1}$ which is an approximate
  estimate of the extent of the linear regime at low redshifts
  \citep{2003ApJ...594..665B}.
\label{figpk}}
\center
\includegraphics[width=10.5cm,angle=-90]{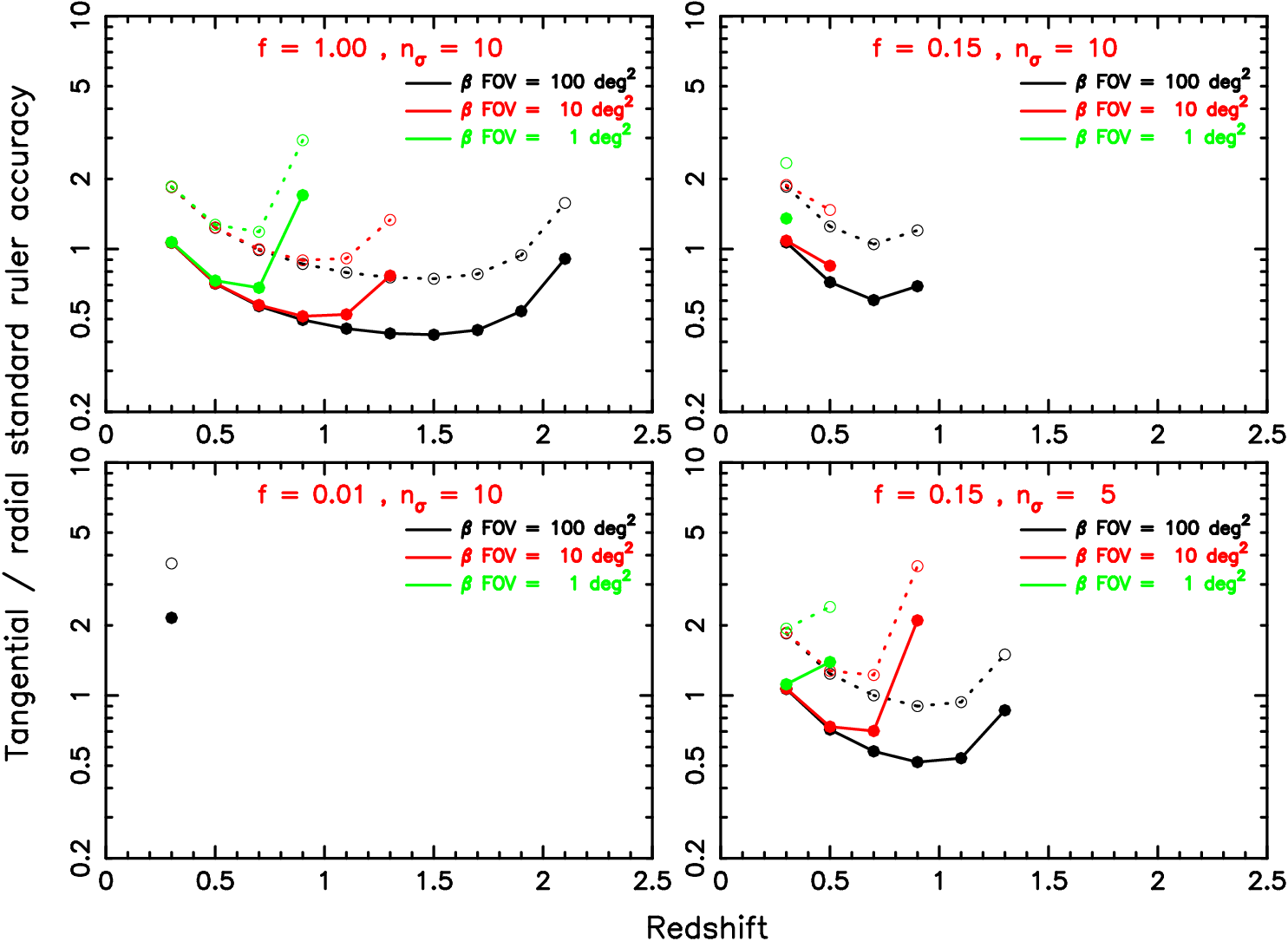}
\caption{Simulated standard ruler accuracies in redshift slices of
  width $\Delta z = 0.2$ for a series of survey parameter sets defined
  by fractional SKA area $f$, threshold for source detection
  $n_\sigma$, and field-of-view $\beta \times FOV$.  The solid lines
  indicate the simulated accuracy of measurement of the tangential
  acoustic scale; the dashed lines display the accuracy of the radial
  scale.
\label{figacc}}
\end{figure*}

We divided each model HI survey into redshift slices of width $\Delta
z = 0.2$ up to the maximum redshift $z_{\rm max}$ defined in
Sec.\ref{seczmax} (rounding up to the next multiple of $\Delta z$).
For each redshift slice we modelled the error in the measurement of
the galaxy power spectrum $P_{\rm HI}(k)$ in each of a series of
Fourier bins of width $\Delta k = 0.01 \, h$ Mpc$^{-1}$:
\begin{equation}
\delta P_{\rm HI}(k,z) = \frac{1}{\sqrt{m}} \left[ 1 + \frac{1}{n(z)
    \, P_{\rm HI}(k,z)} \right]
\label{eq::pHI}
\end{equation}
where $n(z)$ is the HI galaxy number density at redshift $z$, and $m$
is the number of Fourier modes appearing in the bin of width $\Delta
k$:
\begin{equation}
m = \frac{1}{2} \times \frac{V}{(2\pi)^3} \times 4 \pi k^2 \Delta k
\end{equation}
where $V$ is the total survey volume.  The power spectrum errors in
each of the redshift slices are displayed in the panels of
Fig.\ref{figpk} for four example surveys:

\begin{itemize}
\item {\bf Parameter set 1:} $f = 1$, $\beta \, FOV = 10$ deg$^2$,
 $n_\sigma = 10$.
\item {\bf Parameter set 2:} $f = 0.15$, $\beta \, FOV = 10$ deg$^2$,
 $n_\sigma = 10$.
\item {\bf Parameter set 3:} $f = 0.01$, $\beta \, FOV = 10$ deg$^2$,
 $n_\sigma = 10$.
\item {\bf Parameter set 4:} $f = 0.15$, $\beta \, FOV = 10$ deg$^2$,
 $n_\sigma = 5$.
\end{itemize}

In each case the model power spectra and errors have been divided by a
`reference' power spectrum which possesses the same broadband shape
and amplitude as the model power spectrum but has the BAOs erased,
i.e.\ the `no-wiggles' spectrum of \cite{1998ApJ...496..605E}.
Presenting the data in this manner illustrates the approximate
significance with which the baryon oscillations are delineated by
these surveys.  The different redshift slices are offset vertically
for clarity, and data points are plotted up to a maximum value $k =
0.2 \, h$ Mpc$^{-1}$, which is an estimate of the maximum extent of
the linear regime at low redshifts, and which we then conservatively
approximate as being independent of redshift.  We note that an $f = 0.15$
SKA could produce a high-fidelity measurement of baryon oscillations
in each of a series of redshift slices up to $z \approx 0.8$, if
equipped with a field-of-view $\beta \, FOV \gtsimeq 10$ deg$^2$.  A
full SKA could return exquisite measurements of the galaxy power
spectrum up to $z \approx 2$.

\subsection{Fitting formula results}

The BAOs in the galaxy power spectra act as a cosmological standard
ruler which enables the cosmic distance and expansion rate at each
survey redshift slice to be measured with high precision
\citep{2003ApJ...594..665B,2003ApJ...598..720S,2003PhRvD..68f3004H}.
Roughly speaking, oscillations in the tangential direction of the
dataset can be used to accurately map the cosmic distance to each
redshift slice.  Radial oscillations return the Hubble parameter
$H(z)$ at each slice.

\citet{2006MNRAS.365..255B} devised fitting formulae which allow the
survey parameters -- central redshift $z$, volume $V$ and galaxy
number density $n$ -- to be converted into approximate accuracies of
measurement of the tangential and radial baryon oscillation scales.
These formulae are based on the procedure discussed by
\citet{2003ApJ...594..665B} and \citet{2005ApJ...631....1G}, which is
designed to produce robust and conservative estimates of the standard
ruler accuracies (see \citet{2007ApJ...665...14S} for a more recent
fitting formula).

We used the fitting formulae of \citet{2006MNRAS.365..255B} to
determine the tangential and radial standard ruler accuracies in each
of a series of redshift slices of width $\Delta z = 0.2$ for a variety
of surveys varying collecting area $f$, detection
threshold $n_\sigma$ and field-of-view $\beta \, FOV$.  We assume HI
evolution model C.  The results are plotted in Fig.\ref{figacc}.  The
solid and dashed lines plot the tangential and radial accuracy,
respectively, which are approximately equal to the accuracy with which
the quantities $x(z)$ and $H(z)$ may be inferred in each redshift
slice.  This plot serves to emphasize the importance of a large
field-of-view in extending the redshift range over which an HI survey
can map cosmic distances.  Future radio telescopes can produce
per-cent-level measurements of $x(z)$ and $H(z)$ over a range of
redshifts, provided that $\beta \, FOV \, f^2 \ga 1$.

These measurements of cosmic distances and expansion rates may be
readily converted into determinations of the parameters of the dark
energy model.  This is illustrated by Figs.\ref{figomw} and
\ref{figw0wa}, corresponding to two different parameterizations of the
dark energy equation of state: (1) $w(z) = w_{\rm cons}$
(Fig.\ref{figomw}) and (2) $w(z) = w_0 + w_a (1-a)$, where $a=1/(1+z)$
(Fig.\ref{figw0wa}).  We assume Gaussian priors on the other relevant
cosmological parameters -- $\Omega_{\rm m}$ and $h$ -- parameterized
as $\sigma(\Omega_{\rm m} h^2) = 0.004$ and $\sigma(\Omega_{\rm m}) =
0.01$.  This is representative of the combination with measurements of
the Cosmic Microwave Background anisotropies by the forthcoming {\sl
  Planck} satellite.  In detail, we assume that the baryon
oscillations provide a measure of $x(z)/s$ and $H(z)^{-1}/s$, where
$s$ is the sound horizon at recombination, and fold the uncertainties
in $\Omega_{\rm m}$ and $h$ into the error in $s$.

An $f = 0.15$ SKA can produce a strong measurement of a constant dark
energy equation of state ($\sigma(w_{\rm cons}) \la 0.1$) if $\beta \,
FOV \ga 10$ deg$^2$.  A full SKA produces $\sigma(w_{\rm cons})
\approx 0.025$ for $\beta \, FOV = 10$ deg$^2$.  For an evolving
equation of state parameterized by $(w_0, w_a)$, measurements at
higher redshift assume increased importance.  Here, an $f=0.15$ SKA can
produce a good measurement of the evolving term ($\sigma(w_a) \la 1$)
if $\beta \, FOV \ga 50$ deg$^2$ assuming $n_\sigma = 10$.  A more
aggressive threshold $n_\sigma = 5$ doubles the accuracy of the dark
energy measurement in this case.  A full SKA with $\beta \, FOV = 10$
deg$^2$ allows parameter measurements of superb quality: $\sigma(w_0)
\approx 0.05$ and $\sigma(w_a) \approx 0.2$.  In
Sec.\ref{sec:bao_mcmc} we will compare these results with a full MCMC
calculation. 

We also plot in Fig.\ref{fig::newfig}
the accuracies in determining the angular diameter distance and the 
Hubble parameter from some model optical surveys using the fitting
formula as we already described and over plotted onto results 
for radio surveys. This plot gives us an idea of the relative merit of 
optical surveys compared to radio surveys for cosmological 
parameter estimation. A binning of 0.2 in redshift was used and a number
density for optical surveys was assumed to be $nP = 2$ (see Eqn.\ref{eq::pHI}),
the product of the galaxy number density n in the redshift slice and the 
amplitude of the galaxy power spectrum P at $k=0.2$ (i.e. it
is independent of galaxy bias) -- $nP=2$ ensures shot noise is 
less than cosmic variance.

\begin{figure*}
\center
\includegraphics[width=9.5cm,angle=-90]{./figs/hi_omw.ps}
\caption{Simulated dark energy measurements of a constant
  equation-of-state $w(z) = w_{\rm cons}$ for a series of surveys
  defined by fractional SKA sensitivity $f$, threshold for source
  detection $n_\sigma$, and field-of-view $\beta \times FOV$.
  1-$\sigma$ confidence ellipses are plotted in the parameter space of
  $(\Omega_{\rm m}, w_{\rm cons})$.  Marginalized errors are shown for
  the parameter $w_{\rm cons}$.  The analysis uses the measurements of
  $x(z)$ and $H(z)$ using tangential and radial baryon oscillations,
  as displayed in Fig.\ref{figacc}, together with Gaussian priors
  $\sigma(\Omega_{\rm m} h^2) = 0.004$ and $\sigma(\Omega_{\rm m}) =
  0.01$, representative of combining with {\sl Planck} satellite
  observations of the Cosmic Microwave Background.
\label{figomw}}
\center
\includegraphics[width=9.5cm,angle=-90]{./figs/hi_w0wa.ps}
\caption{Simulated dark energy measurements of a redshift-dependent
  equation-of-state $w(z) = w_0 + w_a (1-a)$ for a series of surveys
  defined by fractional SKA sensitivity $f$, threshold for source
  detection $n_\sigma$, and field-of-view $\beta \times FOV$.
  1-$\sigma$ confidence ellipses are plotted in the parameter space of
  $(w_0, w_a)$.  Marginalized errors are shown for $w_0$ and $w_a$.
  The analysis uses the measurements of $x(z)$ and $H(z)$ using
  tangential and radial baryon oscillations, as displayed in
  Fig.\ref{figacc}, together with Gaussian priors $\sigma(\Omega_{\rm
    m} h^2) = 0.004$ and $\sigma(\Omega_{\rm m}) = 0.01$,
  representative of combining with {\sl Planck} satellite observations
  of the Cosmic Microwave Background.
\label{figw0wa}}
\end{figure*}

\begin{figure}
\begin{center}
\centerline{
\includegraphics[width=5.5cm,angle=-90]{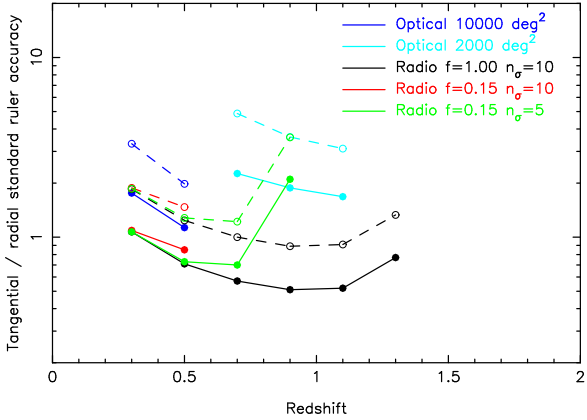}}
\caption{We plot here the accuracies in determining the angular diameter 
distance and the Hubble parameter from some model optical surveys, using 
the fitting formula described and we over-plot results for radio surveys.  
We have kept the binning in dz=0.2 redshift slices The optical surveys we use 
are: 0.2 < z < 0.4, 10000 deg$^2$, 0.4 < z < 0.6, 10000 deg$^2$, 
0.6 < z < 0.8, 2000 deg$^2$, 0.8 < z < 1.0, 2000 deg$^2$ and 1.0 < z < 1.2, 
2000 deg$^2$. $nP = 2$ for all surveys. $nP=2$ refers to the product of the 
galaxy number density n in the redshift slice and the amplitude of the 
galaxy power spectrum P at k=0.2 (i.e. it is independent of galaxy bias) and 
ensures shot noise is significantly less than cosmic variance.
The comparison radio surveys all have $\beta \, FOV = 10$ deg$^2$, 
duration of one year, area of 20,000 deg$^2$. The solid lines (solid circles) 
are angular diameter distances accuracies and the dashed lines (open circles) 
are Hubble parameter accuracies.}
\label{fig::newfig}
\end{center}
\end{figure} 

\section{Simulated dark energy measurements using a full power spectrum analysis}
\label{sec:bao_mcmc}

We now perform simulations using the full power spectrum information
to measure the dark energy model.  The broadband shape of the power
spectrum depends mainly on the quantity $\Omega_{\rm m}h$, and
redshift-space distortions contain information on the large-scale bias
of the galaxy population.  Hence the power spectrum shape helps to
break degeneracies in the cosmological model and sharpen our dark
energy measurements.  We refer the reader to
\citet{2007MNRAS.381.1313A} for a full description of our MCMC
forecast method.

In this Section we assume a survey performed by an SKA with $\beta
\times FOV = 10$ deg$^2$. Therefore, a one-year survey corresponds to
4 hours integration time per pointing. There is a straight-forward
trade-off between collecting area and integration time; when comparing
different collecting areas we specify for reference the integration
time required to achieve an equivalent depth with the full SKA
($f=1$).

\subsection{Effects of survey depth on parameter forecasts}
\label{sub_sec::eff_depth}

\begin{table*}
\begin{center}
\begin{tabular}{|c|c|c|c|c|c|c|c|c|}
\hline
$f$ & 0.15 & 0.25 & 0.35 & 0.50 & 0.70 & 1 & 1.4 & 2 \\
$f= 1$ SKA & 0.1h & 0.25h& 0.5h & 1h  & 2h  & 4h  & 8h  & 16h \\
$z_{\rm max}$ & 0.62 & 0.75 & 0.90 & 1.1  & 1.3  & 1.6  & 1.8  & 2.2 \\
\hline
$\Omega_b$ & 0.0040 & 0.0022 & 0.0014 & 0.0012 & 0.0011 & 0.00090 & 0.00080 & 0.00076 \\
$\Omega_c$ & 0.014 & 0.0078 & 0.0053 & 0.0042 & 0.0038 & 0.0031 & 0.0028 & 0.0021 \\
$w$ & 0.32 & 0.081 & 0.036 &  0.026 &  0.022 &  0.017 & 0.015 & 0.013 \\
$n_s$ & 0.051 & 0.023 & 0.012 & 0.0088 &  0.0073 & 0.0052 & 0.0050 &  0.0047\\
$\sigma_8$ & 0.061 & 0.022 & 0.0087 & 0.0054 & 0.0039 & 0.0028 &  0.0023 &  0.0020\\
$h$ & 0.043 & 0.013 & 0.0058 & 0.0042 & 0.0032  & 0.0025 & 0.0020 & 0.0020\\
\hline
\end{tabular}
\caption[LSS measurements as a function of survey depth.]{The forecast
  errors in cosmological parameters as a function of survey depth.
  The top row indicates the telescope sensitivity (for a 4-hr
  exposure); the second row lists the equivalent exposure time with a
  full SKA ($f=1$).  The third row details the approximate maximum
  redshift achieved, i.e. where $nP = 1$.  
  For small integration times the errors in the
  dark energy equation-of-state $w$ are large because the baryon
  oscillations are not measured accurately.
\label{tab::results_ska1}
}
\end{center}
\end{table*}

We present in this subsection the effect that different collecting
areas (or equivalently different integration times with a full SKA)
have on the accuracy of cosmological parameter measurement for a
$20{,}000$ deg$^2$ survey.  Initially we base the forecasts only on
the SKA large-scale structure data and take a standard $\Lambda$CDM
framework with an extra parameter $w$, a constant equation-of-state of
dark energy.  Our fiducial model is:
\{$\Omega_b$,$\Omega_c$,$w$,$h$,$n_s$,$\sigma_8$,$\tau$\} =
\{0.04,0.26,-1.0,0.72,1.0,0.9,0.09\}. The fiducial bias of each
redshift bin is $b=1$, and we use three redshift bins in the analysis.
We have tested that for a constant-$w$ model there is negligible
benefit from analyzing more redshift bins, although we note that this
may not be the case for a more complex dark energy parameterization.
We present in Table \ref{tab::results_ska1} our forecast errors in
cosmological parameters recovered from the MCMC chains.

Using the large-scale structure data alone, our measurement error in
$w$ degrades sharply at a collecting area below $f \approx 0.35$ (or a
30-min exposure using the full SKA).  This is because below this
approximate threshold we are no longer able to break the degeneracies
between the fitted parameters and we require additional data such as
the CMB.

\begin{figure}
\begin{center}
\centerline{
\includegraphics[width=8.5cm,angle=0]{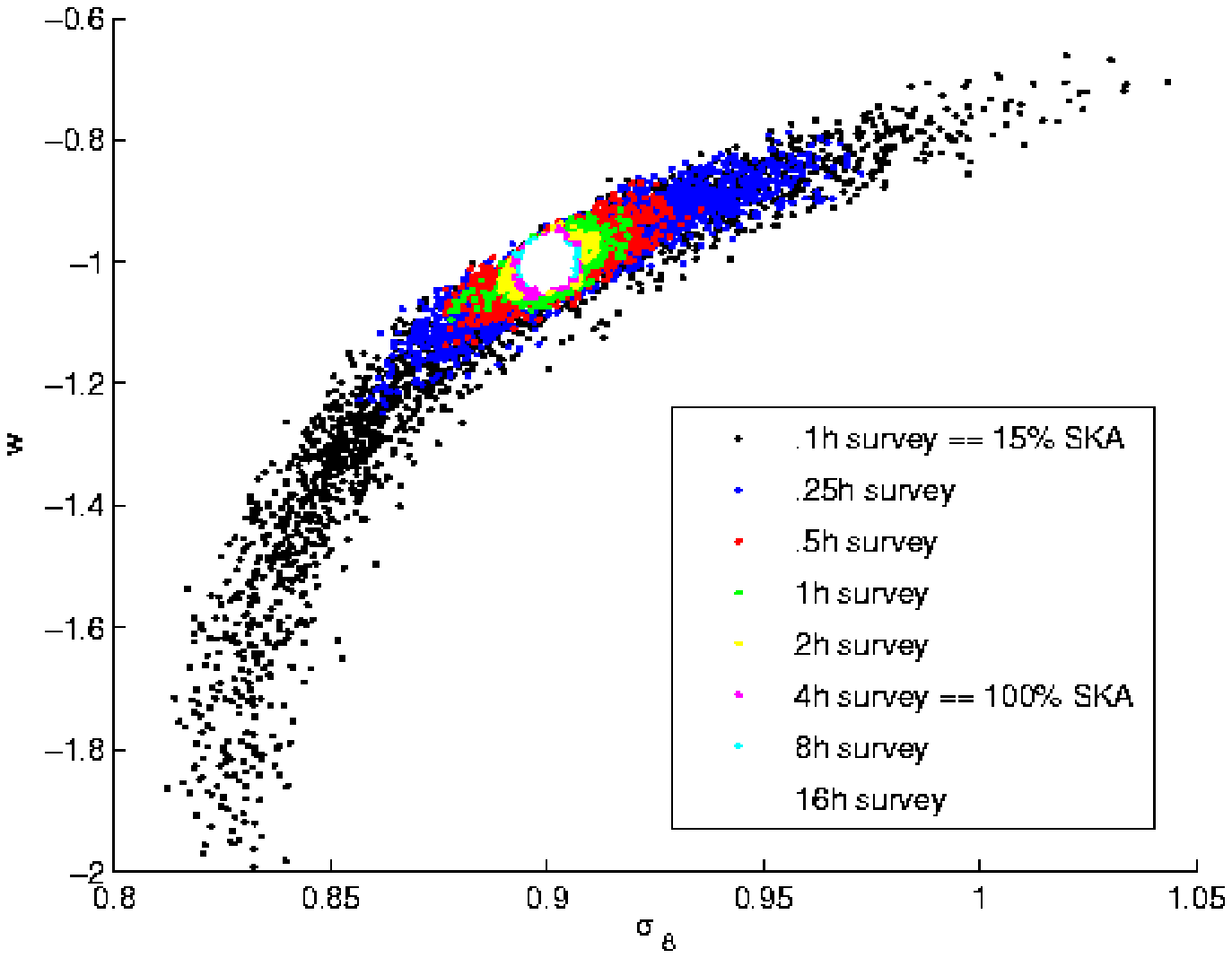}}
\caption[Sensitivity to $w$ given exposure times.]{The posterior in
  the ($\sigma_8$, $w$) plane for full SKA surveys with different
  exposure times. The forecasts improve significantly from 6 mins
  exposure time (equivalent to 4 hrs with an $f=0.15$ SKA) to 4 hrs
  exposure time with the full SKA.  When we reach depths equivalent to
  a 1-hr exposure time with the full SKA, there are sufficient
  galaxies per unit volume and redshift to break the degeneracies and
  produce an accurate measurement of $w$.}
\label{fig::depth}
\end{center}
\end{figure} 

\subsection{A combined experiment}
\label{sec:comb}

\begin{figure*}
\begin{center}
\centerline{
\includegraphics[width=9.5cm,angle=0]{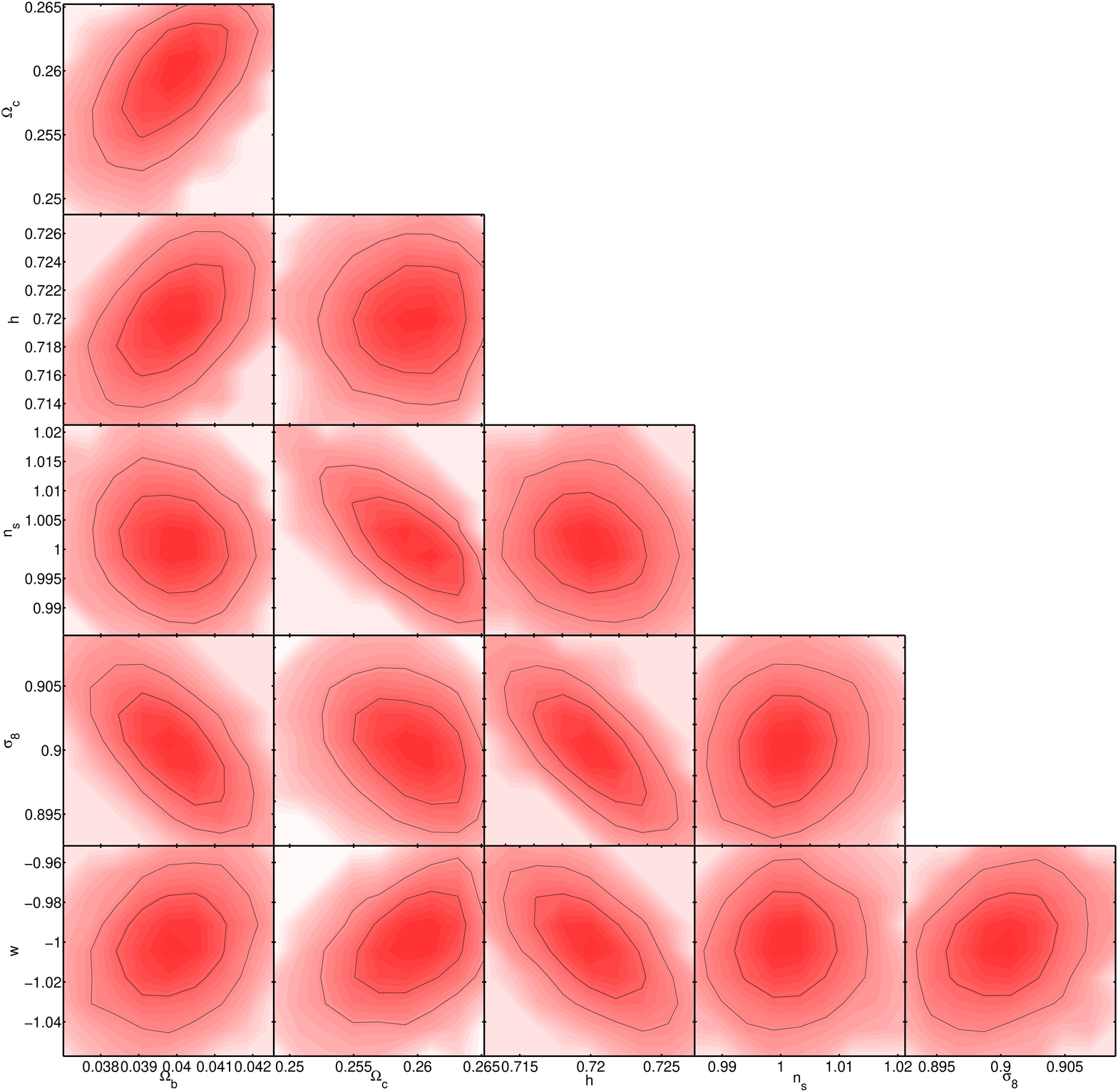}
\includegraphics[width=9.5cm,angle=0]{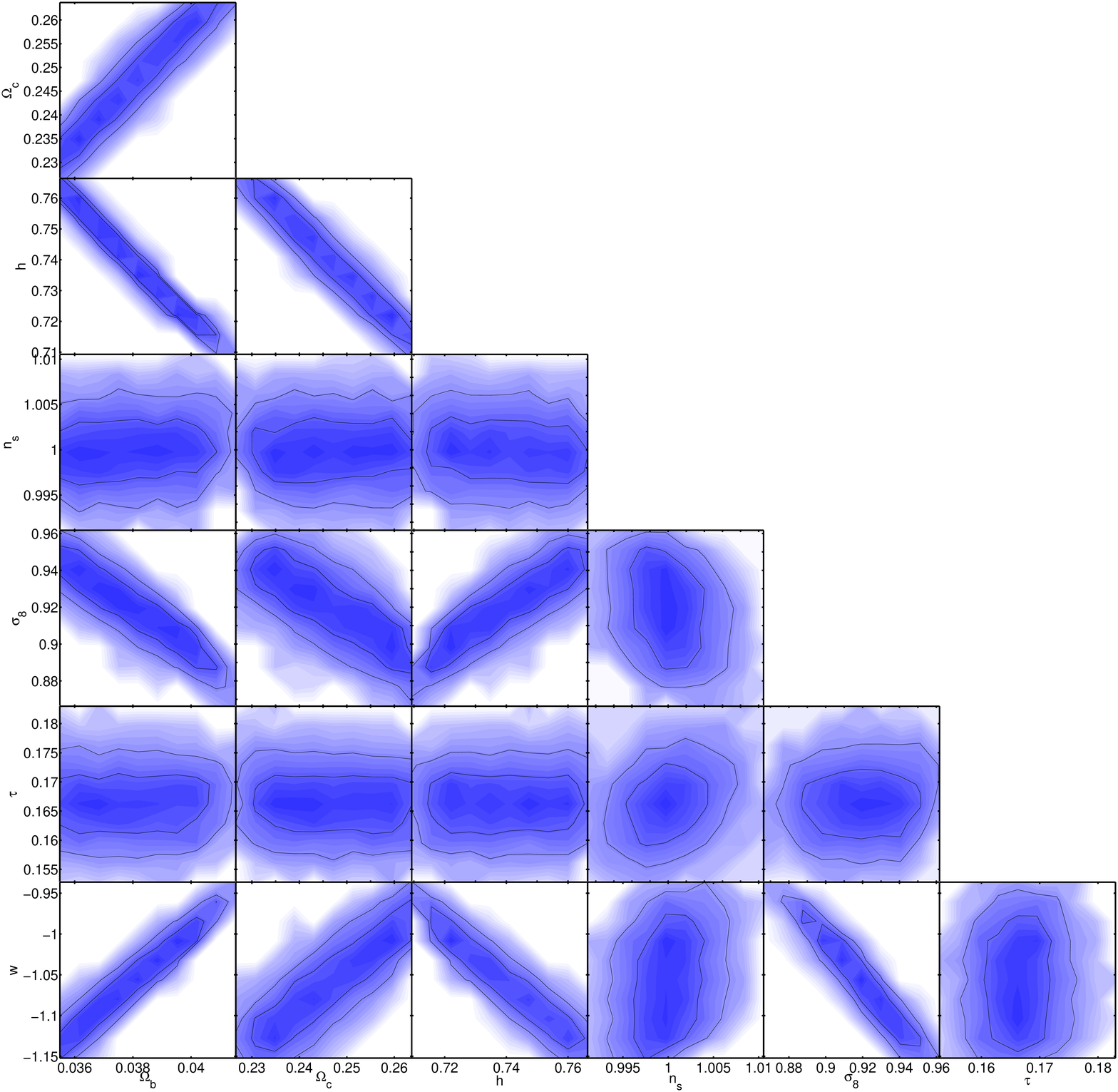}
}
\begin{minipage}[c]{.75\textwidth} 
\centering 
\includegraphics[width=11.5cm,angle=0]{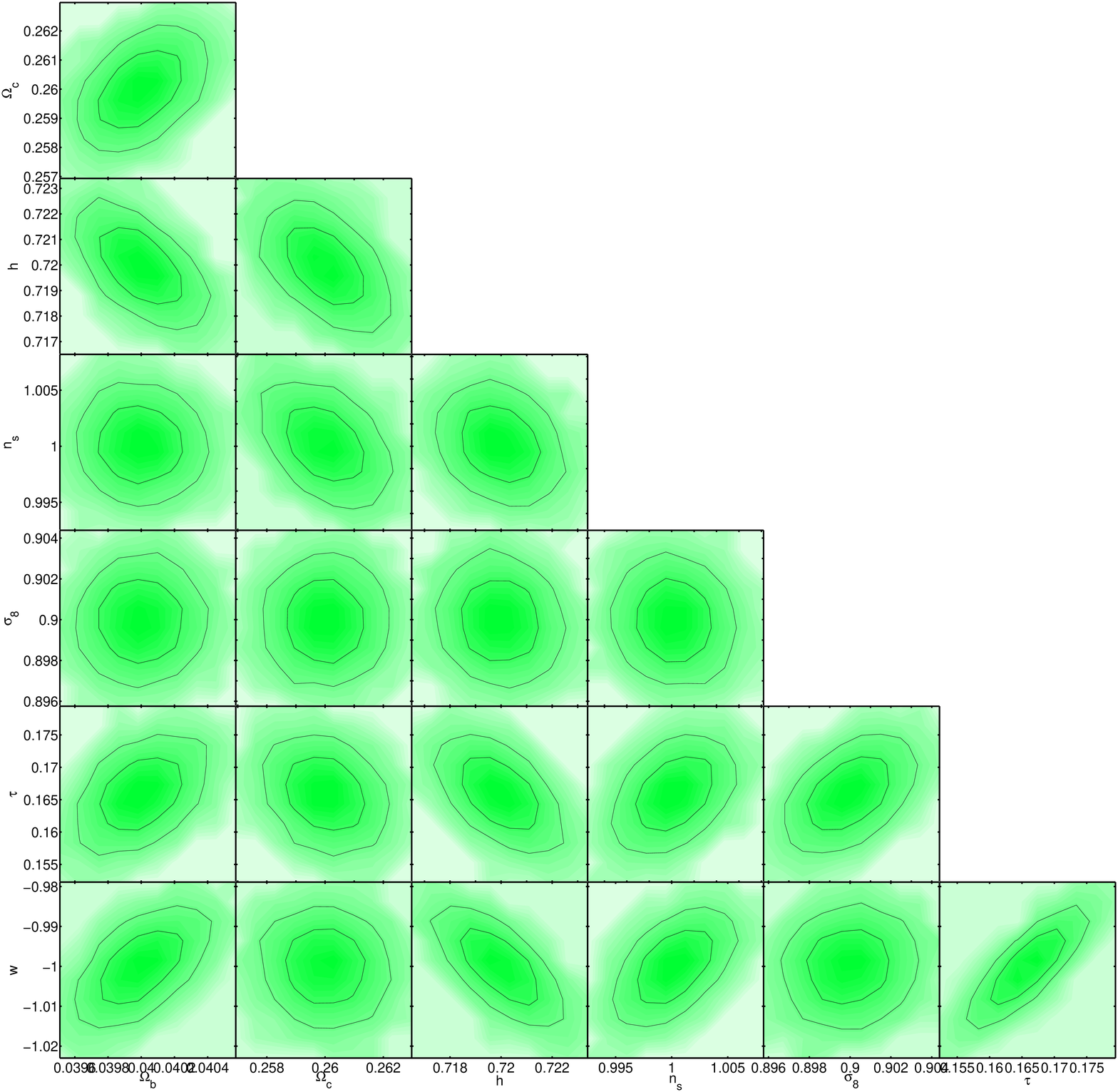}
\end{minipage}
\caption[Measurements of dark energy models from CMB and
  LSS.]{Analysis of the accuracy of a combination of future SKA (red),
  CMB (blue) and combined (green) experiments in measuring
  cosmological parameters. The contours represent one and two sigma
  confidence regions.  The combination of SKA and CMB experiments will
  measure all current cosmological parameters to below the one per
  cent level, including the dark energy parameter $w$.  The full SKA
  will produce a measurement of $w$ with an accuracy close to
  $0.5\%$.}
\label{fig::both_w}
\end{center}
\end{figure*} 

We next forecast the cosmological parameter measurements achieved by
combining the SKA large-scale structure survey with the CMB ({\sl
  Planck}) dataset using the method of \cite{2007MNRAS.381.1313A}.  We
examine how an $f=0.15$ SKA survey and an $f=1$ SKA survey would benefit
from the inclusion of CMB data.  In order to combine the two datasets
we add the log(likelihood) distributions assuming the datasets are
uncorrelated.

\begin{figure}
\begin{center}
\centerline{
\includegraphics[width=9.5cm,angle=0]{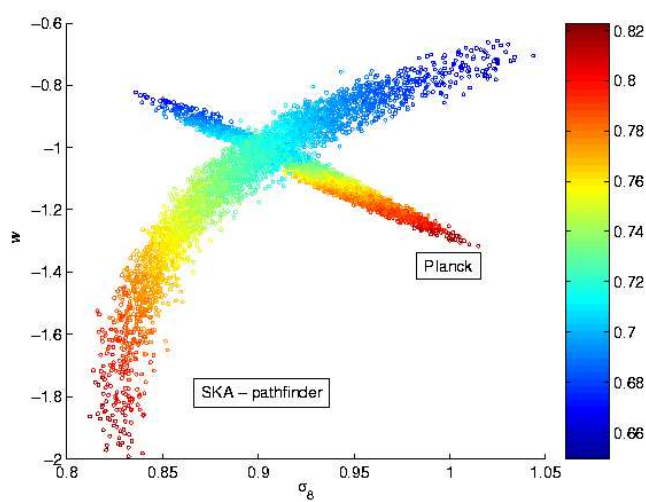}}
\caption[Degeneracies between $w$ and $\sigma_8$.]{We illustrate one
  of the main degeneracies resolved when an HI survey dataset obtained
  with the $f=0.15$ SKA is combined with a CMB ({\sl Planck}) dataset.
  The degeneracy between $\sigma_8$ and $w$ is broken and allows for a
  very good measurement of the dark energy parameter, even with a
  relatively shallow survey covering 20{,}000 ${\rm deg}^2$. The
  samples are for independent MCMC chains for an $f=0.15$ SKA and for {\sl
    Planck}. They are colour-coded according to the value of $h$, as
  shown in the side-bar.}
\label{fig::both_w15}
\end{center}
\end{figure}

We plot in Fig.\ref{fig::both_w} the parameter measurements found for
the combination of full SKA survey and CMB data.  There are no
significant degeneracies in any of the parameter pairs, as opposed to
(for example) the $w-\sigma_8$ plane plotted in Fig.\ref{fig::depth}
from the SKA survey alone (see also Fig.\ref{fig::both_w15}).  The CMB
provides an accurate amplitude of fluctuations at the last scattering
surface, together with a measurement of the angular diameter distance
to this surface, adding complementary information to the SKA
dataset. In Table \ref{tab::results_ska_pl1} we can see that there is
a significant improvement in the parameter errors compared to those
found in Table \ref{tab::results_ska1}.  The accuracy of measurement
of the $w$ parameter is below 1 per cent.

We also conclude from the simulations that most of the constraining
power with regard to the parameter $n_s$ originates from the CMB, and
the LSS dataset adds little.  This is because a high-resolution CMB
experiment with a large multipole range (such as {\sl Planck}) will
measure CMB fluctuations across a large range in spatial scales.  This
picture could change if one were able to model non-linearities in the
large-scale structure including a reliable biasing model at small
scales.  We surmise that the CMB data will be very useful in
constraining some parameters, e.g.\ $n_s$, whilst LSS surveys will
play an increasing role in the determination of other parameters, such
as $\sigma_8$ and $w$ (although the capacity to measure $\sigma_8$
rests on a good model for the galaxy bias from redshift-space
distortions).

We plot in Fig.\ref{fig::both_w15} the MCMC samples obtained for an
$f= 0.15$ `phase I' SKA combined with the future CMB data.  In the
absence of systematic errors this would yield errors in $w$ of
approximately 1 per cent.  This forecast worsens by a factor of 2 if
we assume that redshift-space distortions cannot be used to determine
the galaxy bias factor, as discussed in the next Section.

\subsection{Normalization factors}
\label{ssec::norm_factors}

\begin{figure}
\begin{center}
\centerline{
  \includegraphics[width=9.5cm,angle=0]{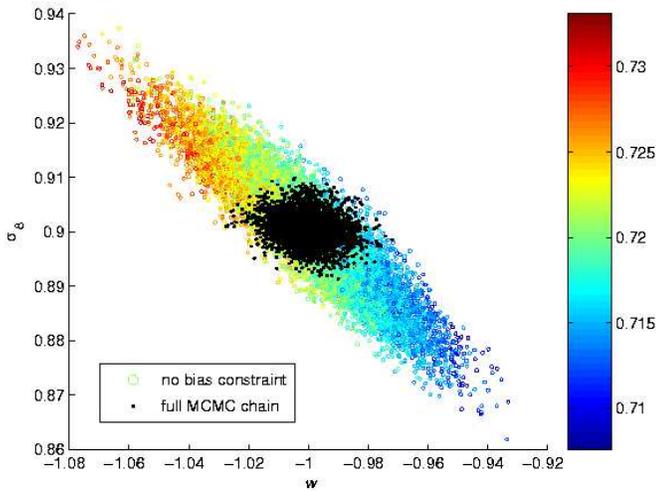}}
\caption[Information content in geometry and growth for an $f=0.15$ SKA +
  {\sl Planck} survey.]{We demonstrate how the accuracy in measuring
  the value of the dark energy parameter $w$ worsens if we cannot
  recover the galaxy bias information.  In this case the distribution
  of MCMC chains changes from the black points to the coloured points,
  a degeneracy between $\sigma_8$ and $w$ opens up, and the error in
  $w$ degrades from $1\%$ to $2.5\%$.  The MCMC samples without bias
  information are colour-coded according to the value of $h$, as shown
  in the side-bar.}
\label{fig::bias_no_bias}
\end{center}
\end{figure} 

The results in Sec.\ref{sec:bao_mcmc} rest on the assumption that
redshift-space distortions can constrain the galaxy bias factor via
large-scale bulk flows
\citep{2008Natur.451..541G,2008arXiv0808.0003P}.  This information can
also be provided by other probes which we have not used here.
Firstly, the analysis of the small-scale clustering in the framework
of the galaxy halo model makes predictions for the large-scale bias
and redshift-space effects.  Secondly, higher-order clustering
statistics yield information on the bias of galaxies
\citep{2002MNRAS.335..432V}.  Reasonably accurate knowledge of the
bias parameter(s) may be gleaned from these tools.  The
scale-dependence of bias may be tested by selecting galaxies with
different clustering properties located in the same comoving volume
\citep{2006astro.ph.11178C}, with the aim of de-coupling the
cosmological signal from the galaxy bias.

On the other hand, we should be cautious in making these assumptions,
since it is not yet clear that the desired level of systematic error
control can be achieved.  In order to investigate the importance of
the additional bias information we performed a simulation considering
only the isotropized power spectrum obtained with an $f=0.15$ SKA survey in
conjunction with the {\sl Planck} CMB data, thereby removing the
information associated with the redshift-space distortions.

We find that the parameter which suffers most from the loss of the
bias information is $\sigma_8$, which is unsurprising given that it is
directly related to the height of the power spectrum.  In
Fig.\ref{fig::bias_no_bias} we plot results from simulations, both
with and without bias measurements from redshift-space distortions.
We find that without the bias information the accuracy of measurement
of $w$ is reduced from the one per cent level to roughly the two and a
half per cent level, and there remains a residual $\sigma_8$-$w$
correlation.

\begin{table}
\begin{center}
\begin{tabular}{|c|c|c|c|}
\hline
Parameter & $f=0.15$ SKA (1) & $f=0.15$ SKA (2) & $f=1$ SKA  \\
\hline
$\Omega_b$ & 0.00024 & 0.00044 & 0.00016 \\
$\Omega_c$ & 0.0016 & 0.0021 & 0.00086\\
$w$ & 0.010 & 0.022 &  0.0062\\
$n_s$ &  0.0024 & 0.0028 & 0.0021\\
$\sigma_8$ & 0.0026 & 0.010 & 0.0013\\
$h$ & 0.0019 & 0.0035 & 0.0010\\
$\tau$ & 0.0042 & 0.0041 & 0.0036 \\
\hline
\end{tabular}
\caption[SKA + {\sl Planck} results.]{The forecast errors in
  cosmological parameters from simulations of an SKA HI survey plus
  CMB ({\sl Planck}) data. The first column of results is for a
  `phase-I' ($f=0.15$) SKA and corresponds to a full analysis including
  the redshift-space distortions effect. The second column assumes
  that this information is unreliable and marginalizes over the
  large-scale anisotropic power. The final column compares these
  results to a full SKA survey including redshift-space distortions.
\label{tab::results_ska_pl1}
}
\end{center}
\end{table}

\subsection{Comparison with BAO-only results}

We can compare these MCMC calculations with our more approximate
results presented in Sec.\ref{sec:bao_fit}.  In the approximation of
Sec.\ref{sec:bao_fit} we only utilize the distance information encoded
in the BAOs, not the full power spectrum shape.  In addition we
incorporate the CMB data only as `representative' priors in the values
of $\Omega_{\rm m} h^2$ and $\Omega_{\rm m}$ rather than by
calculating the full parameter degeneracies, and we do not use
information encoded in the growth of structure.

In Sec.\ref{sec:bao_fit} we forecast an error $\sigma(w) = 0.026$ for
an HI survey with the full SKA with $\beta \, FOV = 10$ deg$^2$.  The
error determination in Table \ref{tab::results_ska_pl1} is $\approx 4$
times smaller.  We suggest that approximately half of this improved
performance is the result of utilizing the growth factor measurements,
and the remainder is the result of using the shape of the power
spectrum to measure the values of $\Omega_{\rm m}$ and $h$ more
accurately hence sharpening our capacity to utilize the BAO distance
information.

\section{Remarks on systematic errors in future radio surveys}
\label{sec:chal}

The forecasts in this paper assume that the HI galaxy clustering
measurements are not compromised by systematic errors from spurious
sources or artifacts in the data.  We present results for 10-$\sigma$
and 5-$\sigma$ source catalogues; however, there is a wealth of extra
information contained in the lower-sigma peaks in our derived radio
images, and our analysis is conservative.  In principle it might be
possible to analyze the raw intensity data cube without the
intermediate step of extracting a catalogue and thereby greatly
enhance the maximum redshift and number density of our HI galaxy
survey \citep{2008PhRvL.100i1303C}.

A concern with this approach is our ability to quantify accurately the
power spectrum of the noise.  Given the dynamic range limitations of
radio interferometry, the noise will be highly variable across the sky
and depend on a host of subtle calibration issues; it is certainly 
likely to be enhanced in the vicinity of bright radio continuum sources
imprinting characteristic scales in the measurement.  The
imprint of this variation will require very careful modelling.  It is
also possible that, in contrast to CMB observations, the noise in the
HI data cube may be non-Gaussian.  A detailed investigation of these
issues is beyond the scope of this paper.

\begin{figure}
\begin{center}
\center
\includegraphics[width=9.5cm,angle=0]{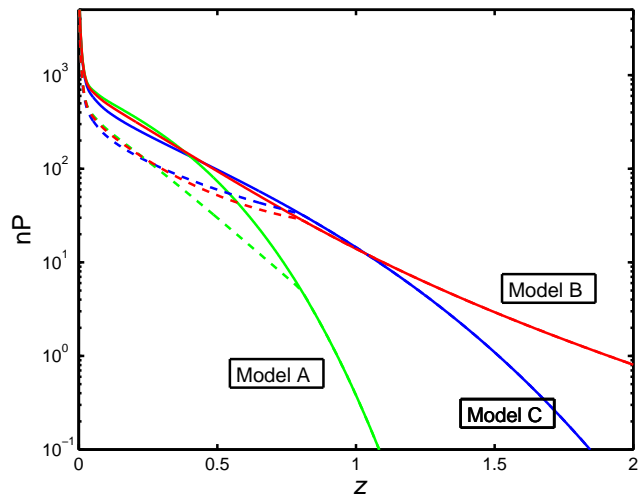}
\caption{The solid lines display the variation of the value of $nP$
  with redshift for the 3 different HI evolution models.  For all
  models the value of $nP$ remains above $1$ for $z < 1$, implying
  that we can measure the galaxy power spectrum for many sub-samples
  and hence constrain the systematic effects of galaxy bias.  The
  dashed lines illustrate the degradation which occurs when the
  telescope sensitivity worsens as the frequency squared in the high
  frequency range, i.e.\ at low redshift.}
\label{fig::np}
\end{center}
\end{figure}

Another source of systematic error in the galaxy power spectrum is the
uncertainty in the galaxy bias factor.  This can be investigated by
measuring the clustering properties of galaxy sub-samples (split for
example by HI mass) across which the imprint of cosmological
parameters should be identical.  This programme can be carried out
effectively for HI surveys owing to the large number density of
sources.  This is illustrated by Fig.\ref{fig::np} -- the solid lines
show the value of the quantity $nP$ for different models, assuming a
1-yr SKA survey with $\beta \, FOV \sim 10$, where $n$ is the galaxy
number density and $P$ is the power spectrum amplitude.  The product
$nP > 10$ over a large redshift range, implying that we can measure a
power spectrum without serious shot noise error for more than $\sim
10$ subsets of galaxies. An aggressive (few $\sigma$) detection threshold
may require more complicated treatments of galaxy bias.
Low-threshold catalogues suffer heavily from Eddington bias effects
that (considering for simplicity one fixed redshift) will yield a
spuriously wide range of intrinsic
HI mass (and hence galaxy bias) at the fixed detection HI line
detection threshold.

Because this survey is `over-sampled' at low redshifts it is possible
to reduce significantly the performance of the telescope without
compromising much of the cosmological science.  In Fig.\ref{fig::np}
we consider the case where the sensitivity decreases with the square of the
frequency above 800 MHz; the result is the dashed line and this could yield
important design considerations: for
example a quadratic
drop-off in the sensitivity of aperture arrays arises naturally if
critical sampling of the wavefront (having antenna elements spaced at
least every half wavelength) becomes relaxed below a wavelength, say
that corresponding to 800 GHz, meaning that at these higher
frequencies the
array becomes sparse rather than close packed \citep{SKAmemo87}.  Gaining a
clearer idea of the frequency ranges where noise can be compromised
depends on a better measurement of the HI mass function at $z \sim 1$,
emphasizing the importance of the stacking experiments mentioned in
Sec.\ref{sec:rad_vs_op}.

In our baseline model the full SKA can survey the $z < 2$ Universe on
the timescale of a year, based on technologies such as a phased array
or dishes with focal-plane phased arrays in order to achieve the
required fields of view.  At lower frequencies, sparsely-sampled
dipoles may be the technology of choice.  In this case
 the scaling of field-of-view with
frequency can be set to the user requirements provided that the
computing power is available to process the required number of beams.
Our assumed $(1+z)^2$ scaling of the $FOV$ is just one choice, that
needs to be optimized with detailed simulations.  At the lowest
frequencies it may be desirable to have the physical collecting area
rising as a sharp function of wavelength to ensure a mapping speed
that is constant in redshift.  For a sparse array this is naturally
achieved via the $\lambda^2$ variation in the collecting area of
antennas. However it may still be necessary to build physically more
antennas in this frequency range, or to use the processing system to shape
how the FOV varies with frequency, and hence redshift, by, for
example,
forming more beams at the lower frequencies to yield greater
instantaneous sky coverage. This is subject to the fundamental
limitation of the finite sky area observable at moderate zenith angle
from an aperture array station.

\section{Conclusions}
\label{sec:conc}

We have forecast the cosmological parameter measurements that could
result from a combination of future HI-emission galaxy surveys with
the SKA and CMB ({\sl Planck}) data.  The main degeneracies currently
present in cosmological data would be broken by such new surveys.
Current parameters would be measured to below the $1\%$ level and
other parameters such as neutrino masses will be accessible.

We find that an HI survey with a telescope possessing $f=0.15$ of the
SKA sensitivity, together with a future CMB experiment, will be able
to determine the dark energy equation of state with an accuracy close
to one per cent.  We have examined the origin of this information and
found that the robust signature of `geometry' (the projection of the
BAOs) is producing significant power.  The remaining information
originates from measurements of the growth of cosmic structure that
can be inferred by looking at redshift-space distortions at large
scales and thereby breaking the degeneracy with galaxy bias.

We caution that the evolutionary model considered here adopts a slightly
stronger evolution in the cosmic mass density of HI (see Fig. 2) than is
inferred from the most recent studies of HI absorption 
\citep{2008arXiv0810.3236P}, so the observing times required may be 
underestimated by a small
factor (assuming most of the number counts are coming from the exponential 
tail of the mass function near the break, which should be the case in the 
high redshift region; in order to have $\sim$ 1.5 times more sources, 
one would need to detect sources $\sim$ 
1.7 times the mass, which would require 
$\sim$ 2.8 the exposure time roughly). We
emphasise the importance of stacking experiments in Sec.\ref{sec:rad_vs_op}.

A survey with the full SKA would be able to produce a measurement of
the dark energy parameter with an accuracy close to half a per cent.
In this case the addition of the CMB data does not produce such a
substantial increase in performance.  However, the SKA and CMB data
remain complementary in the sense that the galaxy survey data is
better able to measure the height of the power spectrum (hence
$\sigma_8$) and the CMB data is most powerful for determining the
spectral index of scalar fluctuations ($n_{\rm s}$).

We conclude that a radio telescope with a sensitivity between $f=0.15$ and
$f=1$ of the SKA would be a cosmological tool capable of
transformational science, particularly in the field of dark energy.

\section*{Acknowledgments}

We thank Kathryn Nicklin and Danail Obreschkow for practical help with
the work of this paper and for useful discussions.  FBA acknowledges
support from the Leverhulme Foundation via an Early Careers
Fellowship. This work has been undertaken as part of the Square
Kilometre Array Design Study (SKADS) financed by the European
Commission.

\bibliographystyle{mn2e} 
\bibliography{./reference/aamnem99.bib,./reference/ref_data_base.bib}

\end{document}